\documentclass[pra,twocolumn,superscriptaddress,floatfix,10pt]{revtex4-1}
\usepackage{graphicx,bm}
\usepackage{amsmath}
\usepackage{amsfonts}
\usepackage{amssymb}
\usepackage{braket}
\usepackage{dsfont}
\usepackage{bm}

\usepackage{color}

\usepackage{xcolor}
\usepackage[%
colorlinks=true,
urlcolor=blue,
linkcolor=blue,
citecolor=blue
]{hyperref}

\newcommand{\bk}{\boldsymbol{k}}

\begin{document}

	\title{Glide symmetry breaking and Ising criticality in the quasi-1D magnet CoNb$_2$O$_6$}
	\author{Michele Fava}
	\affiliation{Rudolf Peierls Centre for Theoretical Physics, Department of Physics, University of Oxford, Oxford OX1 3PU, United Kingdom}
	\author{Radu Coldea}
	\affiliation{Clarendon Laboratory, Department of Physics, University of Oxford, Oxford OX1 3PU, United Kingdom}
	\author{S.A. Parameswaran}
	\affiliation{Rudolf Peierls Centre for Theoretical Physics, Department of Physics, University of Oxford, Oxford OX1 3PU, United Kingdom}

	\begin{abstract}
		We construct a microscopic spin-exchange Hamiltonian for the quasi-1D Ising magnet CoNb$_2$O$_6$ that captures detailed and hitherto-unexplained aspects of its dynamic spin structure factor. We perform a symmetry analysis that recalls that an individual Ising chain in this material is  buckled, with two sites in each unit cell related by a glide symmetry. Combining this with numerical simulations benchmarked against neutron scattering experiments, we argue that the single-chain Hamiltonian contains a staggered spin-exchange term. We further argue that the transverse-field-tuned quantum critical point in CoNb$_2$O$_6$ corresponds to breaking this glide symmetry, rather than an on-site Ising symmetry as previously believed. This  gives a  unified microscopic explanation of  the dispersion of confined states in the ordered phase and `quasiparticle breakdown' in the polarized phase at high transverse field.
	\end{abstract}

	\maketitle

	Magnetic materials with a large mismatch of exchange strengths along different  crystalline axes can often 
	be understood from a one-dimensional (1D) starting point. In this paper, we focus on a celebrated  example of such a quasi-1D magnet~\cite{Vasiliev2018}, CoNb$_2$O$_6$~\cite{PhysRevB.60.3331,Heid1995,MAARTENSE197793,SCHARF1979121}, usually viewed as a system of weakly-coupled  ferromagnetic Ising chains~\cite{Lee2010}. Several theoretical predictions rooted in this perspective that leverage techniques ranging from integrability and conformal field theory (CFT)~\cite{doi:10.1142/S0217751X8900176X, PhysRevD.18.1259, DELFINO1996469, DELFINO1995724, 2006hep.th...12304F} to matrix-product state numerical methods~\cite{PhysRevB.83.020407} have been verified via neutron scattering experiments. Especially striking among these are studies of the transverse-field-tuned quantum critical point (QCP)~\cite{PhysRevLett.112.137403,PhysRevX.4.031008,liang2015heat,Bach2019}, considered a canonical example of the Ising universality class~\cite{sachdev_2011}. However, many detailed  experimental features have resisted a fully microscopic explanation. This is particularly true away from the critical regime, where perturbations to the simplest  Ising description  play a significant role.

\begin{figure}
	\begin{center}
	    \includegraphics[width=.95\linewidth]{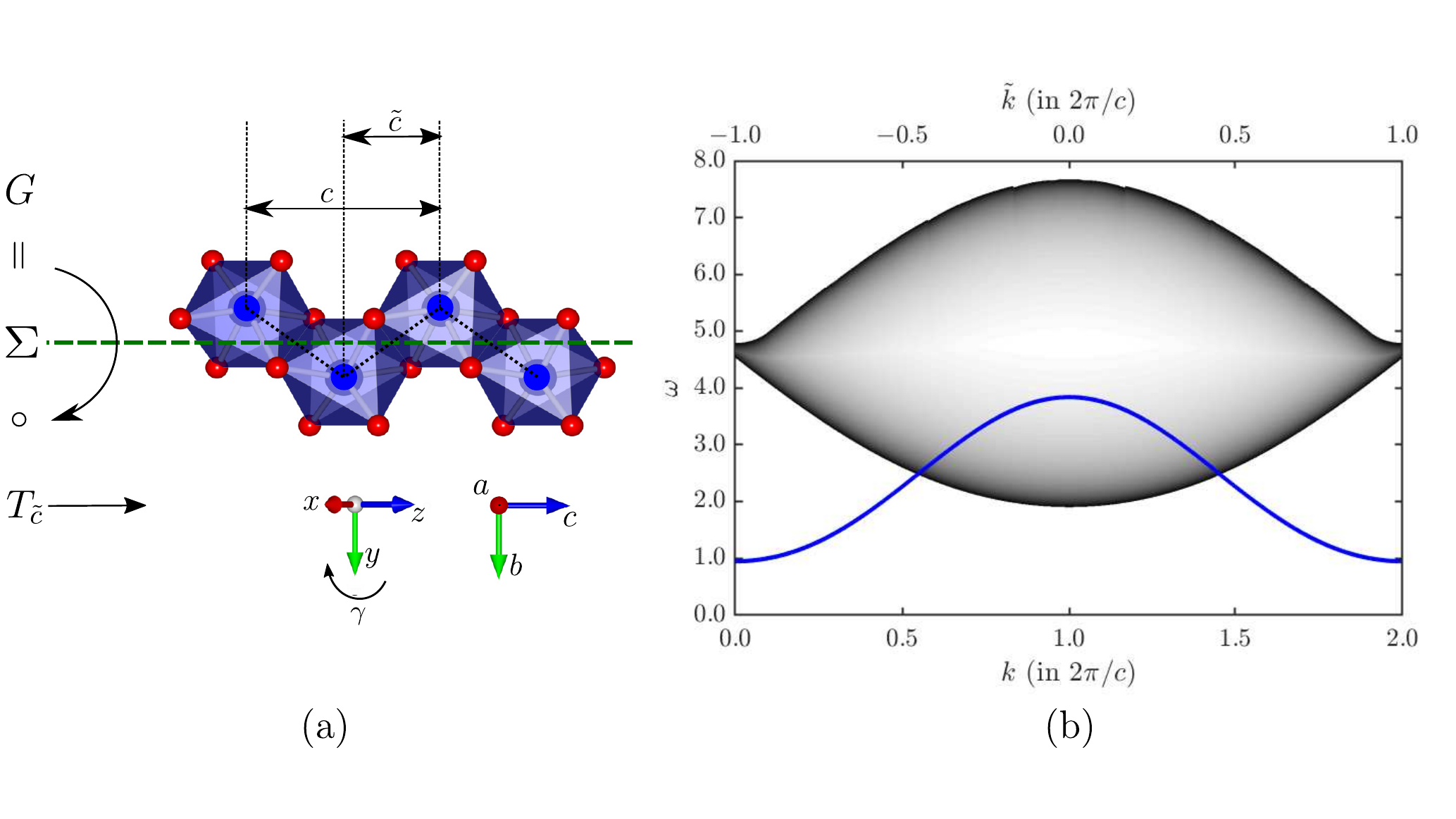}
	\end{center}
		
		\caption{(a) Individual Ising chain in CoNb$_2$O$_6$. The zig-zag structure gives rise to a glide reflection symmetry, corresponding to an $ac$-plane reflection $\Sigma$ followed by a half-lattice translation $T_{\tilde{c}}= T_{c/2}$.(b) Excitation spectrum in the transverse-field polarized quantum paramagnetic phase, showing the band of single spin-flip  QPs (blue solid line)  and the 2QP continuum (shaded region; darker shading indicates higher density of states). {The $x$-axis shows both  momentum $k$ (bottom) and glide eigenvalue $\tilde{k}$ (top).}  The position of the 2QP continuum relative to the 1QP band is controlled by conservation of $\tilde{k}$, rather than $k$, as detailed in the \emph{Quasiparticle breakdown} section.
		}
		\label{fig:glide}
	\end{figure}

	Here, we revisit the models used to describe CoNb$_2$O$_6$, paying attention to the fact that it is only a {\it quasi}-1D system. We use a combination of symmetry analysis, time-dependent density-matrix renormalization group (tDMRG) simulations, and exact diagonalization studies to construct a microscopic 1D model, compute its dynamical spin structure factor (DSF), and compare against that measured by inelastic neutron scattering (INS) experiments.
	By exploring various symmetry-allowed exchange terms beyond the dominant Ising coupling, we find that  
	the origin of various hitherto-unexplained features of the DSF may be traced to a {\it single} previously-ignored contribution: namely, a staggered nearest-neighbor exchange between $y$- and $z$-axis spin components ($z$ is  the Ising axis).
	Its admissibility originates in a subtle and oft-overlooked feature of CoNb$_2$O$_6$, namely that {the magnetic Co$^{2+}$ ions are arranged in zig-zag chains oriented along the $c$-axis as shown in Fig~\ref{fig:glide}a, with the primitive unit cell for one chain containing two Co$^{2+}$ sites with staggered displacements along the $b$-axis.}
	In other words, it relies on the fact that the chain is embedded in a 3D crystal,  leading to distinct symmetry considerations than in pure 1D. We show that this term controls both (i)  the dispersion of confined bound states of two domain wall (DW) excitations~\cite{Coldea2010,PhysRevB.83.020407,Rutkevich_2010} in the spontaneously ordered phase for zero and weak transverse fields, previously only captured phenomenologically; and (ii) quasi-particle (QP) breakdown, a phenomenon observed~\cite{PhysRevB.90.174406} in the opposite limit when a strong transverse field drives the system into a {polarized} quantum paramagnet. 
	
	The staggered coupling we consider explicitly breaks both Ising and translational symmetry. Nevertheless, we show that it  may be reconciled both with the host of results predicated on the existence of an Ising QCP in CoNb$_2$O$_6$ and with the absence of Brillouin zone (BZ) halving in the DSF throughout the field-polarized phase.
	This is because the chain retains a glide symmetry composed of translation by half a lattice spacing ($\tilde{c}=c/2$ in Fig 1(a)) combined with a reflection; the ordered phase breaks this symmetry. A unitary transformation maps the symmetries of our model to those of an Ising {\it antiferromagnet} in a transverse field, and the glide-breaking transition to the breaking of discrete translational symmetry in that model, known to be in the Ising universality class~\cite{PhysRevB.66.075128,PhysRevB.68.214406,NETO20131}. 
	The transformed Hamiltonian has a single-site unit cell, and the transformation effectively ``unfolds'' the BZ probed by DSF into one twice as large (size $4\pi/c$) --- i.e., the same as that observed in experiments.  Unit cell doubling is manifest in the DSF only upon {\it breaking} the glide symmetry, allowing  scattering to directly probe the glide-breaking order parameter. We  show that glide symmetry provides a natural kinematic interpretation of quasiparticle breakdown {in the high-field-polarized phase}: the decay processes that drive it are constrained by glide selection rules rather than by momentum conservation.   
	This scenario provides a natural explanation of several important features of experimental INS data.
	Our work gives a fully microscopic understanding of Ising criticality, confinement, and QP breakdown in CoNb$_2$O$_6$, providing a platform for further studies, and illustrates the subtle interplay of spatial and on-site symmetries in chain- and layer- compounds with non-symmorphic space groups.

\section{Symmetries and Microscopic Hamiltonian}

    In CoNb$_2$O$_6$, Co$^{2+}$ ions hosting pseudospin-$1/2$ moments are arranged in zig-zag chains along the $c$-axis, with a dominant ferromagnetic Ising coupling along the chain. The chains  form a triangular lattice  in the $ab$ plane, with nearest-neighbor chains  weakly  coupled antiferromagnetically. The 3D space group is $Pbcn$ (space group no. 60 in the nomenclature of ref.~\onlinecite{cristallography}), which includes a glide plane that intersects each chain perpendicular to its zig-zag plane {[$ac$-plane in Fig.~\ref{fig:glide}(a)]}. Henceforth we focus on a single chain, and denote by  $S_j^\alpha$ the $\alpha$-component of the spin operator acting on the $j$-th site (even and odd sites lie on different sublattices), with  $\alpha\in\{x,y,z\}$, {defined as in Fig.~\ref{fig:glide}(a). The Ising axis $z$ lies in the $ac$-plane at an angle $\gamma=29.6^{\circ}$ to $c$~\cite{Heid1995}, while $b$ is parallel to $y$.} 
    
    We propose that  the one-dimensional physics in CoNb$_2$O$_6$ is well-captured by  the minimal single-chain Hamiltonian
    \begin{eqnarray}
		\label{eq:hamiltonian}
		\mathcal{H} &=& J\sum_j\left[- S_j^z S_{j+1}^z + \lambda_{\text{AF}} S_j^z S_{j+2}^z + h_y S^y_j  + \right. \nonumber\\
		& &	\left. -\lambda_S \left(S_j^x S_{j+1}^x + S_j^y S_{j+1}^y \right) - h_z  S_j^z\right] + \mathcal{H}_{dw}.
	\end{eqnarray}
	This  includes, apart from the leading nearest-neighbor ferroomagnetic Ising exchange coupling, a next-nearest neighbor antiferromagnetic $\lambda_{\text{AF}}$ term~\cite{PhysRevB.83.020407,PhysRevB.90.174406} as well as an effective longitudinal field $J h_z=J_{\text{ic}}^{zz} \sum_j \langle  S_j^z\rangle/L$ accounting for interchain coupling (parameterized by  $J^{zz}_{\text{ic}}$) at the mean-field level. Both of these (as well as $\lambda_S$) are needed to reproduce details of the experimental zero-field DSF~\cite{Coldea2010}. We also include a transverse field $Jh_y = g_b\mu_B B$, where $B$ is an external magnetic field.
	All terms in  \eqref{eq:hamiltonian} have been previously identified, except for the final term
	$\mathcal{H}_{dw}$ 
	which is required to give dynamics to DWs when $h_y=0$; since its magnitude~\cite{Coldea2010,PhysRevB.83.020407} is much larger than that of inter-chain couplings (which are $\sim h_z$~\cite{PhysRevB.90.014418}), it cannot arise primarily from these. A key goal of this work is to identify a microscopic origin for this DW hopping, that was previously only modeled phenomenologically~\cite{PhysRevB.83.020407}.
    
    To identify $\mathcal{H}_{dw}$, we focus on nearest-neighbour couplings, which are likely dominant, and use a symmetry analysis to narrow our search.
    Recalling
    {that each unit cell has two magnetic sites (per chain)}, the nearest-neighbor exchange Hamiltonian takes the form
	\begin{equation}
		\mathcal{H}_{\text{nn}} = \sum_j \sum_{\alpha,\beta} \mathcal{J}^{(1)}_{\alpha,\beta} S_{2j}^\alpha S_{2j+1}^\beta + \mathcal{J}^{(2)}_{\alpha,\beta} S_{2j+1}^\alpha S_{2j+2}^\beta.
	\end{equation}
	Crystal symmetry further constrains $\mathcal{J}^{(n)}_{\alpha,\beta}$. Inversion about the mid-point between consecutive spins requires  $\mathcal{J}^{(n)}_{\alpha,\beta}=\mathcal{J}^{(n)}_{\beta,\alpha}$, while the glide symmetry imposes 
	\begin{equation}
		\mathcal{J}^{(1)}_{\alpha,\beta} = \left\{\begin{array}{cc} - \mathcal{J}^{(2)}_{\alpha,\beta} & \text{if } \beta=y\neq \alpha,\\  \mathcal{J}^{(2)}_{\alpha,\beta}& \text{otherwise}. \end{array} \right.
	\end{equation} 
	The most general exchange tensor satisfying these conditions may be parametrized (in the $xyz$ basis) as
	\begin{equation}
		\label{eq:J-tensors}
		\mathcal{J}^{(m)} = J
		\begin{pmatrix}
		\lambda_S +\lambda_A & (-1)^m\lambda_{xy} & \lambda_{xz} \\
		(-1)^m\lambda_{xy} & \lambda_S -\lambda_A & (-1)^m\lambda_{yz} \\
		\lambda_{xz} & (-1)^m\lambda_{yz} & -1
		\end{pmatrix}
	\end{equation}
	with $m=1,2$ and $J>0$. Taking $z$ along the Ising axis sets $\lambda_{xz}=0$. Since the  nearest-neighbour ferromagnetic coupling $J$ dominates,  we anticipate all $\lambda\lesssim 1$.

	\begin{figure}
	\begin{center}
	    \includegraphics[width=\linewidth]{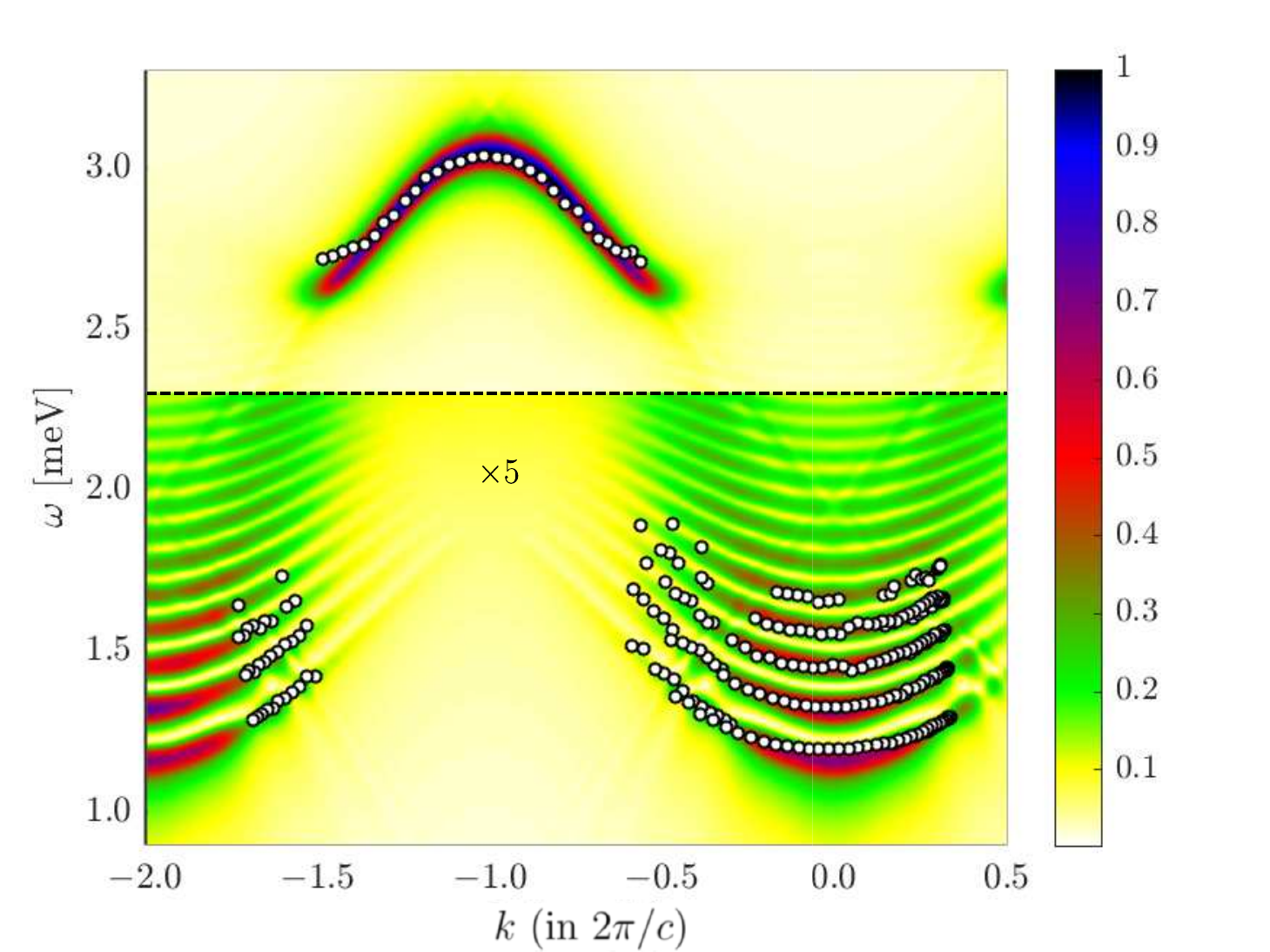}
	\end{center}
		\caption{
			DSF ${\mathcal{S}}^{xx}(\omega,k)$ at $h_y=0$, scaled by a factor of $5$ below the dashed line. Dots denote dispersion data points from the INS data of Ref.~\cite{Coldea2010}. (tDMRG simulations truncated singular values $\lesssim\varepsilon=2\cdot10^{-11}$ and used a Trotter step $\delta t = 2.5\cdot10^{-3}/J$ and   
			broadening $\eta=J/125$~\cite{SM}.)
		}
		\label{fig:hy_0}
	\end{figure}

	\begin{figure*}[t!]
	\begin{center}
	    \includegraphics[width=.95\linewidth]{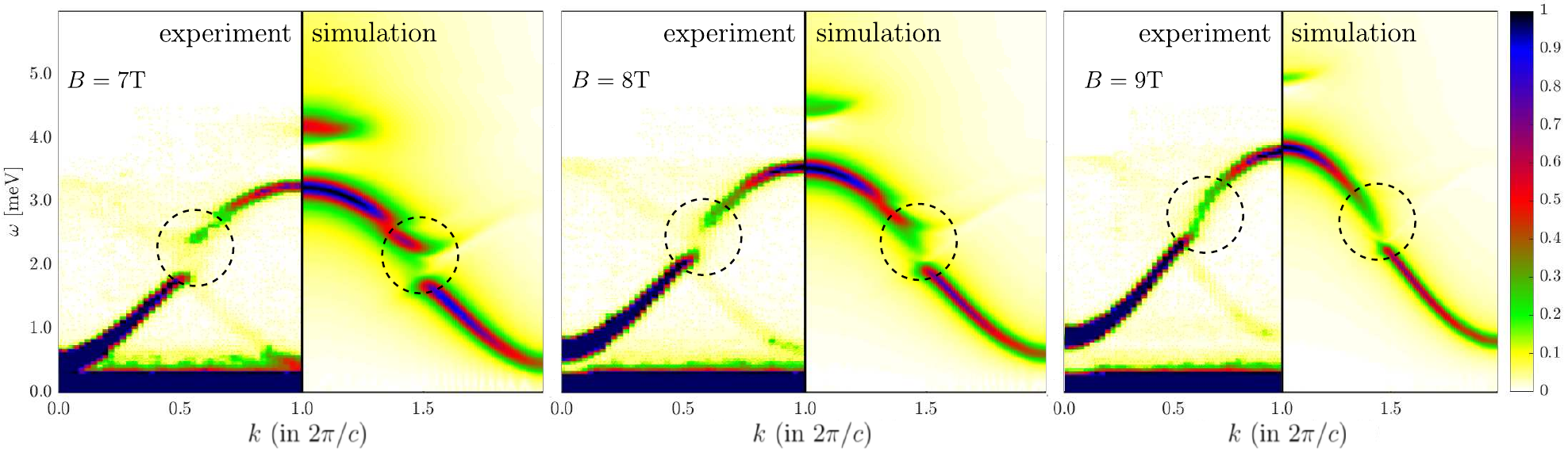}
	\end{center}
		\caption{
			Evolution of DSF in the polarized phase at large transverse field - as obtained from tDMRG simulations of $S_{xx}(\omega, k)$ for the 1D Hamiltonian Eq.~\eqref{eq:hamiltonian} (right image in each pair) - compared with inelastic neutron scattering data from ref.~\cite{PhysRevB.90.174406} (left image in each pair).
			Dashed circles indicate the regions of QP breakdown. (tDMRG parameters: $\delta t = 5\cdot10^{-3}/J$, $\varepsilon=2\cdot10^{-10}$, $\chi_{max}=400$, $\eta=J/60$~\cite{SM}).
		}
		\label{fig:paramagnetic_phase}
	\end{figure*}

	Neutron scattering probes  the DSF, which is related to the spin-spin correlation function of Hamiltonian $\mathcal{H}$ via
	\begin{equation}\label{eq:DSFdef}
	{\mathcal{S}}^{\alpha\beta}(\omega, k) \equiv \frac{1}{L} \int_{-\infty}^\infty d t \sum_{i,j} e^{ik(r_i - r_{j})} e^{i\omega t} \langle S^\alpha_ i(t) S^\beta_{j}\rangle,
	\end{equation}
	where $S^\alpha_j(t) = e^{i\mathcal{H}t} S^\alpha_j e^{-i\mathcal{H}t}$ and $L$ is the total number of sites.  For  transverse ($b$-axis) field $h_y=0$, operators $S^{x,y}_j$  excite a pair of DW excitations of the (Ising) ordered state. These are confined by a longitudinal ($z$-axis) mean field $h_z$.
	Of all the symmetry-allowed nearest-neighbor exchange terms in (4) only $\lambda_{yz}$, $\lambda_A$ and $\lambda_{xy}$ produce DW hopping. We neglect the latter two terms in a first approximation as they create a DW continuum with double the $k$-space periodicity of that seen in experiments. This leads us to consider
	 	\begin{equation}
	\label{eq:DW_hopping}
	\mathcal{H}_{dw} = J\sum_j\lambda_{yz}  (-1)^j \left( S_j^z S_{j+1}^y + S_j^y S_{j+1}^z \right).
	\end{equation}

	We have also verified that to $ O(\lambda_{yz})$, the full Hamiltonian in ~\eqref{eq:hamiltonian} including $\mathcal{H}_{dw}$ reproduces the effective Hamiltonian for DW motion used to parametrize the $h_y=0$ experimental data~\cite{Coldea2010, PhysRevB.83.020407}.
	 While  $\mathcal{H}-\mathcal{H}_{dw}$ is symmetric under translation ${T}_{\tilde{c}}: r_j \mapsto r_j + \tilde{c}$ by a nearest-neighbor spacing $\tilde{c}$,  $\mathcal{H}_{dw}$ only preserves translation ${T}_c$, with $c = 2\tilde{c}$. However, this unit cell doubling is invisible in the DSF  which is consistent with a BZ of size $2\pi/\tilde c$ (see Fig.~\ref{fig:hy_0}). Below, we link this to  a non-symmorphic glide symmetry ${G} \equiv {T}_{\tilde{c}} \circ {\Sigma} $, i.e. a translation by half a unit cell composed with a spin-flip ${\Sigma} = e^{i \pi \sum_{j} S^y_j}$ in the $ab$ plane (consistent with the spatial reflection of a pseudovector spin). First, however, we determine the magnitude of the    couplings in $\mathcal{H}$.

\section{Numerical results}

    We fix the parameters in $\mathcal{H}$ using exact diagonalization on small system sizes with $h_y=0$. 
	Through the fitting procedure described in the supplement~\cite{SM}, we find
	$J =2.7607~\text{meV}$,  $\lambda_{\text{AF}} = 0.1507$,  $\lambda_S = 0.2392$,   $\lambda_{yz} = 0.1647$, and 
	$J_{\text{ic}}^{zz} = 0.0312~\text{meV}$.
	For the Hamiltonian thus obtained we compute the DSF
	for an effectively infinite system using tDMRG~\cite{SM}.
	Our numerical results match  the experimental data well (Fig.~\ref{fig:hy_0}). 
	We also compute the DSF in the high-field quantum paramagnetic  regime, achieved for  sufficiently strong transverse  (i.e., $b$-axis) field. We set $h_z=0$, consistent with the fact that the inter-chain mean field vanishes when $\langle S^z_j\rangle=0$. {To match simulations with data, we estimated $g_b\simeq3.100$ by fitting the experimental dispersion at $B=7$T~\cite{SM}. A direct comparison of our results against data from~\cite{PhysRevB.90.174406} (Fig.~\ref{fig:paramagnetic_phase}) shows  excellent agreement, including features associated with ``quasiparticle breakdown'' --- i.e., the apparent break in the dispersion of the QP band, traditionally understood as a decay of a QP    
	as it enters the two-QP continuum, that occurs for sufficiently strong coupling (see, e.g.~\cite{PhysRevB.90.174406}). We now rationalize these results in terms of  symmetries of $\mathcal{H}$.
	
\section{Ising criticality, glide symmetry, and BZ unfolding}

    Hamiltonian \eqref{eq:hamiltonian} has neither translational symmetry by a nearest-neighbor spacing (${T}_{\tilde{c}}$) nor on-site Ising symmetry (generated by
	${\Sigma}$)  as neither commutes with $\mathcal{H}_{dw}$. However, it preserves their {product}: glide symmetry ${G}$.

	We now consider the unitary transformation ${U} = {U}^{-1}= e^{i\pi \sum_j S^y_{2j}}$ which reverses the Ising axis on alternate sites of the chain. It is straightforward to see that the invariance of $\tilde{\mathcal{H}} = {U} \mathcal{H} {U}^{-1}$  under $T_{\tilde c}$ is equivalent to that of $\mathcal{H}$ under ${G}$. The transformation flips the sign of the nearest-neighbor $zz$ and $xx$ couplings and staggers the $h_z$ term, while leaving the $h_y$ term  unchanged. Crucially  $\tilde{\mathcal{H}}_{dw}$ is no longer staggered, and hence preserves ${T}_{\tilde{c}}$, but continues to  break global Ising symmetry $S^z_j \mapsto -S^z_j$. Thus, $\tilde{\mathcal{H}}$ describes a translationally-invariant  Ising antiferromagnet (AF) in a uniform transverse field $h_y$, with additional terms that break global Ising symmetry, augmented with
	a field $h_z$ that couples to the AF order parameter field $m_s = \sum_j (-1)^j \langle S^z_j\rangle/L$. For $h_z=0$, $\tilde{\mathcal{H}}$ continues to have a transition in the Ising universality class, since the AF order parameter spontaneously breaks the lattice symmetry~\cite{PhysRevB.66.075128,PhysRevB.68.214406} (see also~\cite{NETO20131}). Reversing the unitary transformation, we see that this corresponds to the spontaneous breaking of the glide symmetry of $\mathcal{H}$. In the ordered phase  $h_z\neq 0$ due to the inter-chain mean field, and couples to the order parameter field $m_s$. Therefore, the near-critical ordered phase is described by the Ising CFT perturbed by the magnetization operator ---  
	precisely that for which $E_8$-symmetry-related bound states were predicted~\cite{doi:10.1142/S0217751X8900176X} and experimentally identified~\cite{Coldea2010}. Therefore, although  our revised model \eqref{eq:hamiltonian}  associates the Ising criticality of CoNb$_2$O$_6$ with the spontaneous breaking of glide symmetry rather than the on-site Ising symmetry, it remains consistent with previously-reported experiments. 

	The unitary transformation also allows us to view scattering experiments as probing the DSF of $\tilde{\mathcal{H}}$ (up to a $k$-space shift). To see this, observe that the DSF of  $S^\mu_k = \sum_j e^{ikj\tilde c} S^\mu_j$ under the dynamics generated by $\mathcal{H}$ is equal to the DSF of $\tilde{S}^\mu_k = U S^\mu_k U^{-1}$ under the dynamics generated by $\tilde{\mathcal{H}}$, as can be seen by inserting  $U^{-1}U=\mathds{1}$ in \eqref{eq:DSFdef}. Either by studying the commutation relations of $\tilde S^\mu_k$ with $T_{\tilde c}$ or directly by inspecting $\tilde S^\mu_k = \sum_j e^{ikj\tilde c} U S^\mu_j U^{-1}$, we see that $S^\mu_k$ changes the momentum as $\tilde k = k+\delta_\mu$ where $\delta_{x,z} = \pi/\tilde c$, and $\delta_y=0$. Consequently, since  $\tilde{\mathcal{H}}$ has a unit cell of length $\tilde c$, the DSF will be $2\pi/\tilde c$-periodic as long as $G$ is unbroken (i.e. in the high-field paramagnetic phase).
	In the ordered phase, where $G$ is broken, we expect that the DSF is only $2\pi/c$-periodic (i.e. sees a smaller BZ). This is  also corroborated by INS data Fig.~\ref{fig:ordered_phase}. Note however that exactly at $h_y=0$ we recover $2\pi/\tilde c$-periodicity of the DSF  as evident in Fig.~\ref{fig:hy_0}, since $\mathcal{H}$ has an extra glide symmetry given by $G' =  T_{\tilde c} \circ e^{i\pi \sum_j S_j^z}$, which is explicitly broken for  $h_y\neq 0$.
	[The same conclusions follow from the commutators of $G$ with $S^\mu_k$, as the operators $G$ and $T_{\tilde c}$ are related by the unfolding  $U$~\cite{SM}.]
	
    {Note that, rigorously speaking, this unfolding is possible only if different chains are decoupled. The presence of inter-chain couplings in the actual material means that the DSF will have small corrections not accounted for in our argument. However, we expect these corrections to be weak, since the inter-chain couplings are two orders of magnitude weaker than the intra-chain couplings~\cite{PhysRevB.90.014418}, as can also be inferred from our estimate of $J^{zz}_{\text{ic}}$.}
    
	We now use this BZ unfolding perspective to interpret experiments on $\mathrm{CoNb_2O_6}$.
	Scattering non-polarized neutrons from a sample should yield superposition of the DSF of $\tilde H$ (from the $y$ component) and the same DSF shifted in momentum $k$ by $\pi/\tilde c$ (from the $x$ and $z$ components). However in the high-field polarized phase $\langle S^y_j \rangle$ is near saturation, thus largely suppressing the inelastic $b$ component of the DSF. Hence, the dominant signal observed in this phase is the one at $k = \tilde k + \pi/\tilde c$; shifting the glide-labeled spectrum by $\pi/\tilde{c}$ thus reproduces the measurements. 
	{[A weaker-intensity ``shadow mode''   shifted by a wave-vector $\pi/\tilde c$ visible in the experimental data is due to the fact that the scattering wavevector is not aligned with the $c$-axis, but has a non-zero component in the $b$-direction, in addition to the component $k$ along $c$~\cite{PhysRevB.90.014418,SM}. We note that alternative explanations of this mode that invoke inter-chain couplings can be ruled out due to the negligible magnitude of the latter.]}

	\begin{figure}
	    \begin{center}
	        \includegraphics[width=\linewidth]{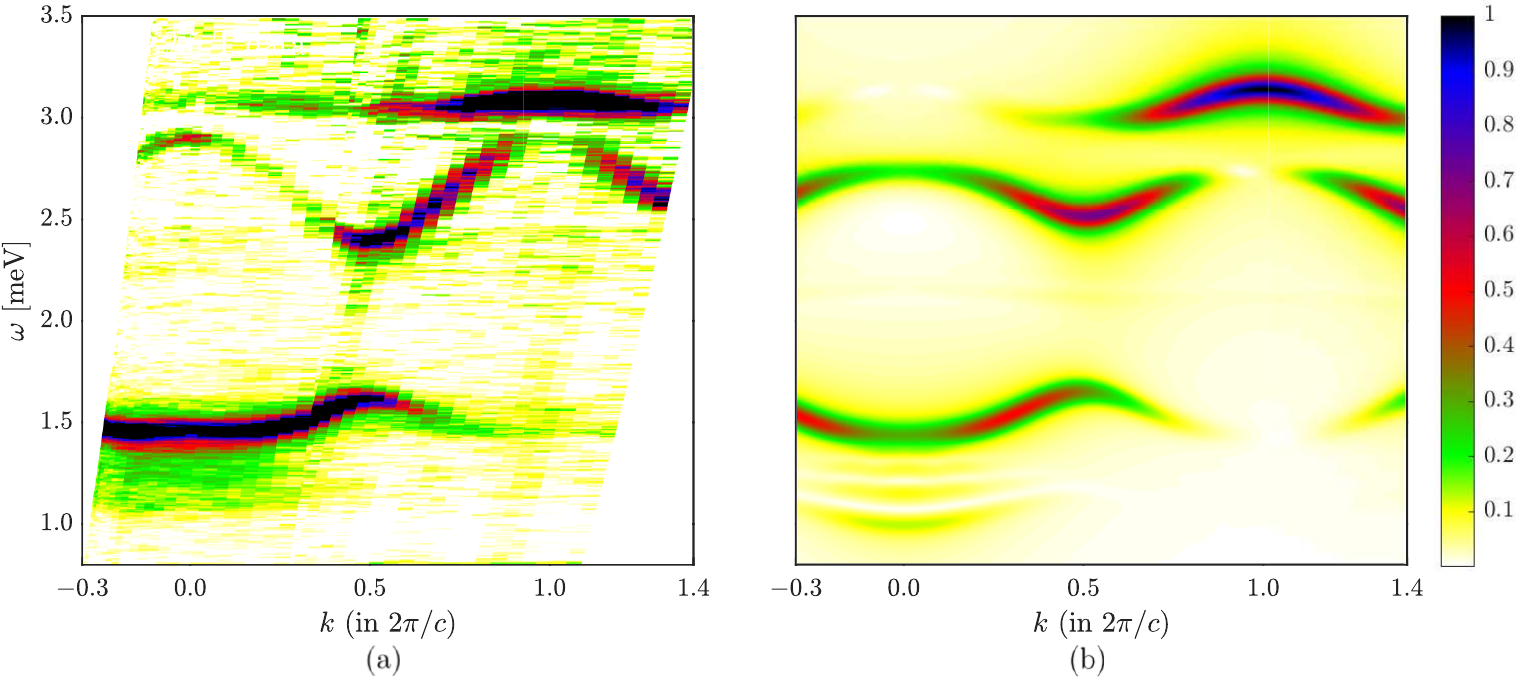}
	    \end{center}
		\caption{$(\text{a})$ DSF measured by INS experiments in the ordered phased at intermediate transverse field ($2.5$ T) collected using the same experimental setup as in the field-dependent INS measurements reported in~\cite{Coldea2010}. Sharp modes show the reduced periodicity of the structural $2\pi/c$ BZ, in contrast to the $4\pi/c$ periodicity in the ordered phase at zero field (Fig.~\ref{fig:hy_0}) and the high-field paramagnetic phase (Fig.~\ref{fig:paramagnetic_phase}). This is tied to the breaking of glide symmetry $G$ throughout  the ordered phase except at $h_y=0$, which enjoys an extra glide symmetry $G'$.
		$(\text{b})$ 
		DSF $S^{xx} (k,\omega)$ calculated via tDMRG ($\delta t = 2.5\cdot10^{-3}/J$, $\varepsilon=2\cdot10^{-11}$, $\eta=J/100$~\cite{SM}).  Key features of the spectrum in $(\text{a})$ including all dispersion shapes are qualitatively well-reproduced, justifying the minimal Hamiltonian ~\eqref{eq:hamiltonian}. More precise quantitative agreement requires  fitting data to the tDMRG simulations of ~\eqref{eq:hamiltonian} rather than a perturbative low-energy model~\cite{SM}. This  is very computationally demanding and beyond the scope of this work.}
		\label{fig:ordered_phase}
	\end{figure}
	
\section{Quasiparticle breakdown}

    The elementary excitations of the high-field phase are dressed spin-flips relative to the $b$-axis field.
	For $\lambda \neq0$, these are interacting quasiparticles with rich physics. As seen in experiments and reproduced by our model  (Fig.~\ref{fig:paramagnetic_phase}), they exhibit the striking feature of quasiparticle breakdown.
	Glide symmetry  provides a natural resolution of why  the decay of a single QP into the 2-QP continuum that causes this effect is limited to a narrow, field-dependent range of momenta, 
	(Fig.~\ref{fig:paramagnetic_phase}). First, matching the $k$ values with their respective $\tilde k$ as described above, we  find that the minimum (maximum) of the single-QP dispersion $\omega_{\text{QP}}(\tilde{k})$ is at $\tilde k=\pi/\tilde c$ ($\tilde k=0$) as shown in Fig.~\ref{fig:glide}. We stress that only $\tilde k$ is a good quantum number for $\mathcal{H}$, rather than the experimental wavevector  $k$. With this in mind, the states that form part of the two-QP continuum in the $(\tilde{k},\omega)$ plane are those satisfying $\tilde{k}= \tilde{k}_1 +\tilde{k}_2$, $\omega = \omega_{\text{QP}}(\tilde{k}_1)+\omega_{\text{QP}}(\tilde{k}_2)$. From this, we see that the entire upper section of the single-QP band is immersed in the continuum (Fig.~\ref{fig:glide}(b)).
	We emphasize that this is sharply distinct~\cite{SM} from  approaches that do not incorporate the glide symmetry and the corresponding $\pi/\tilde c$ momentum shifts (compare Fig.~\ref{fig:glide}(b) to Figs.~2,~6 of \cite{PhysRevB.90.174406}).	
	
	Now, the QP-continuum matrix elements
	are at most $O(\lambda_{yz})$, as for $\lambda_{yz}=0$ the decay would be forbidden by $\mathbb{Z}_2$ (Ising) symmetry. Using Fermi's Golden rule and neglecting to first approximation the dependence of matrix elements on momenta,  the  QP decay rate at wave vector $k$ may be estimated as $\Gamma(k) \sim |\lambda_{yz}|^2 \rho_2(\tilde{k},\omega(\tilde{k}))$. Here  $\rho_2(\tilde{k},\omega)$ is the two-QP density of states (DoS), which is large (essentially an $({\omega-\omega_c(k)})^{-1/2}$ singularity possibly renormalized by interactions) 
	near the edges of the 2QP continuum~\cite{Gaveau1995,RevModPhys.85.219,Verresen2019}. Hence, although in much of the 2QP continuum the decay rate is suppressed by $|\lambda_{yz}|^2\ll 1$, the 
	large DoS  near its edge allows  full QP breakdown. This explains the relatively narrow region in momentum space where breakdown is visible in Fig.~\ref{fig:paramagnetic_phase}. In future high-resolution numerical and experimental studies, it may be interesting to probe the detailed transfer of spectral weight between the 1QP band and 2QP continuum for  signatures of ``avoided quasiparticle decay''~\cite{Verresen2019}.
	
\section{Discussion}

    We have constructed a  microscopic spin-exchange Hamiltonian to describe the 1D physics of  {CoNb$_2$O$_6$}, based on a symmetry analysis of its 3D space group. We found quantitative agreement between  simulations of the model and INS experiments in very different regimes, indicating that the proposed model realistically captures single-chain physics.
	
	A crucial departure from previous studies lies in the symmetries of our model, which identifies a two-site unit cell and explicitly breaks the on-site Ising $\mathbb{Z}_2$ symmetry. However, it retains a glide symmetry inherited from the 3D space group. We showed through an explicit unfolding transformation that the glide symmetry leads to a larger BZ for INS than that expected from the size of the unit cell. This transformation also shows that the model is consistent with previous reports of Ising criticality in {CoNb$_2$O$_6$}, if the ordering is linked to the breaking of glide symmetry rather than on-site $\mathbb{Z}_2$.
	
	{From a more fundamental perspective, the model presented in this paper is a microscopic justification --- based on the actual symmetries of the material --- that a quantum critical phase transition with gap closing and re-opening can exist in CoNb$_2$O$_6$. Experimentally there are indeed abundant indications of a proximity to a 1D QCP in the Ising universality class in this material. However, gap closing has not yet been reported experimentally and careful inspection of the crystal structure shows in fact that the material does not microscopically feature an exact purely magnetic Ising symmetry. Therefore there is no \textit{a priori} reason why the field-induced transition should be continuous and in the Ising universality class. Our work provides a clear answer to such questions: our model shows how the microscopic symmetries of CoNb$_2$O$_6$ admit a true QCP in the Ising universality class, with an order parameter that transforms under both magnetic and space-group symmetries.}
	
	Our work  reveals how  subtle aspects of crystal symmetry intertwine with the rich physics of quantum criticality, and provides a unified picture of spontaneous ordering, confinement, and quasiparticle breakdown in a canonical Ising-chain system. It raises further interesting questions as to how symmetry considerations impact the rich 3D phase structure of 	{CoNb$_2$O$_6$}~\cite{doi:10.1063/1.3562516,Lee2010,PhysRevB.60.3331,Heid1995,SCHARF1979121}. Similar ideas are likely relevant to other chain (e.g. BaCoV$_2$O$_8$~\cite{PhysRevLett.114.017201, Faure2018}, Sr$_2$V$_2$O$_8$~\cite{PhysRevB.91.140404}, SrDy$_2$O$_4$~\cite{PhysRevB.89.224511, PhysRevB.93.060404} or azurite Cu$_3\left(\text{CO}_3\right)_2\left(\text{OH}\right)_2$~\cite{PhysRevB.83.104401,PhysRevLett.106.217201}) and layer compounds with non-symmorphic space groups~\cite{SM}.

\section{Materials and methods}

    \subsection{Fit of parameters for $h_y=0$}
	
	In order to fix the $5$ parameters ($J$, $\lambda_{\text{AF}}$, $\lambda_S$, $\lambda_{yz}$ and $J^{zz}_{\text{ic}}$) of the Hamiltonian for $h_y=0$, we resort to exact diagonalization on a small system of size $L=12$. We checked that the dispersions of the $2$ lowest confined bound states, unlike those of  higher ones, are not strongly affected by finite size effects at $L=12$. Furthermore, their dispersion can be quickly accessed by targeting the low-energy subspace using the Lanczos algorithm. We thus fitted parameters to minimize the square difference of the energy dispersion as obtained in two different ways: (i) experimentally, from INS~\cite{Coldea2010} and (ii) numerically,  by interpolating exact-diagonalization results on an $L=12$ site system, restricting our attention to the lowest two modes.
	
	In order to avoid overfitting, we constrained the parameters to reproduce (at first order in perturbation theory) the dispersion of the kinetic mode as parametrized in Ref.~\cite{Coldea2010}. This fixes
	\begin{align}
		J\lambda_{S} &= 0.6605 \mbox{ meV}\\
		(1-\lambda_{\text{AF}})J &= 2.3447 \mbox{ meV}		
	\end{align}
	thus leaving only three free parameters to be fitted.
	
	\subsection{Details of the tDMRG simultations}
	
	Working at $T=0$, we computed the matrix-product state (MPS) approximation of the ground state $\ket{0}$ of $\mathcal{H}$ on a chain of $L$ sites using DMRG. From this, the state $\ket{\psi}=S_{L/2}^\beta\ket{0}$ can be computed. We then time-evolve the state, viz. $\ket{\psi(t)}=\exp(-iHt)\ket{\psi}$, through tDMRG, up to some maximum time $t_{max}$. The DSF can then be be approximated by~\cite{Paeckel:2019yjf}
	\begin{widetext}
	\begin{align}
		S^{\alpha\beta}(\omega, k) = 
		E\left[2 \Re \left(\int_0^{t_{max}} d t \sum_{i} e^{ik(r_i - r_{L/2})} e^{i(\omega+\omega_0) t} \Braket{0 | S^\alpha_ i |\psi(t)} \mathcal{W}(t)\right)\right].\nonumber
	\end{align}
	\end{widetext}
	Here, we define $E[f](q)=f(q)+f(-q)$, which, exploiting the inversion symmetry about bond centres, is equivalent to averaging over the position of $j$. Finally $\mathcal{W}(t)$ is a windowing function smoothly suppressing contributions at larger $t$, in such a way that the truncation at $t=t_{max}$ does not produce oscillations in the Fourier transform (this requires $\mathcal{W}(t_{max})\ll1$). In this paper, we use a Gaussian windowing $\mathcal{W}(t) \varpropto \exp\left[-(\eta t)^2 /2\right]$~\cite{PhysRevB.77.134437}. Due to this choice the DSF obtained from the computation can be viewed as the convolution of the exact DSF with a Gaussian of width $\eta$, broadening spectral lines.
	
	For a fixed broadening $\eta$, there are $3$ parameters controlling the errors in the computation: the Trotter step $\delta t$ in the time evolution, the singular value cutoff $\varepsilon$ in the SVD and an hard maximum $\chi_{\text{max}}$ of the maximum singular values we retain in the SVD. The values of these parameters are indicated in the caption of the corresponding figure. When $\chi_{\text{max}}$ is not explicitly reported, it is meant that an hard-cutoff was unnecessary as the bond dimension growth was mild.
	Exact results are recovered as $\varepsilon\to0$, $\chi\to\infty$ and $\delta t \to 0$.
	The convergence analysis for the results we report is presented in Ref.~\cite{SM}.  Numerical simulations were performed using the ITensor Library~\cite{ITensor}.
	
	\subsection{Data Sharing}
	All numerical data of the theoretical calculations including all code used in the theoretical analysis and the experimental data points in Fig.~\ref{fig:hy_0} are available from the Oxford University Research Archive~\cite{ORA}. The experimental data in Fig.~\ref{fig:ordered_phase} are available from the corresponding author on request. The experimental data in Figs.~\ref{fig:hy_0} and~\ref{fig:paramagnetic_phase} are adapted/reproduced here from refs.~\onlinecite{Coldea2010, PhysRevB.90.174406}.

\begin{acknowledgements}
	We  thank  Sarang  Gopalakrishnan, Fabian  Essler, Frank  Pollmann, John  Chalker, and  Alexander  Chernyshev  for  useful  discussions.  We  also  thank  Ruben  Verresen  for  useful  comments  on  the manuscript. We acknowledge support from the  the European Research Council under the European Union Horizon 2020 Research and Innovation Programme via Grant Agreements No. 788814-EQFT (RC)  and 804213-TMCS (SAP).
\end{acknowledgements}

\bibliography{IsingGlide.bib}

\clearpage

\onecolumngrid
\newpage
\setcounter{equation}{0}
\renewcommand{\theequation}{S\arabic{equation}}
\setcounter{figure}{0}
\renewcommand{\thefigure}{S\arabic{figure}}

\section*{SUPPLEMENTARY INFORMATION FOR ``Glide symmetry breaking and Ising criticality in the quasi-1d magnet CoNb$_2$O$_6$''}

\renewcommand{\theequation}{S\arabic{equation}}

\renewcommand{\thefigure}{S\arabic{figure}}

\setcounter{figure}{0}

\setcounter{equation}{0} 

\begin{appendix}
	\section{Dispersion of the 2-DW spectra in perturbation theory for different terms}
	\label{app:effective-hamiltonian}
	
	In this appendix we derive an effective Hamiltonian for $\mathcal{H}$ with the addition of other symmetry allowed terms in zero external magnetic field $h_y=0$. To do so, we treat perturbatively all the couplings w.r.t. the ferromagnetic one. We expect this approach to give qualitatively correct results, since all couplings $\lambda$ are at most of order $10^{-1}$.
	In particular, we consider (one at a time) the effect of a non-zero $\lambda_{xy}$, $\lambda_{yz}$ and $\lambda_{A}$ (on top of the other terms in $\mathcal{H}$, Eq.~(5)
	).
	
	$\mathcal{H}_f = -J \sum_j S_j^z S_{j+1}^z$ splits the Hilbert space in highly degenerate multiplets, the energy of which is set by the number of domain walls. At first order in perturbation theory, we neglect the mixing between different multiplets and project the couplings of the Hamiltonian into a given multiplet. Since we are interested in eigenstates that are connected to the ground states through a single spin-flip, as these are the ones giving a dominant contribution to the DSF, we focus on the multiplet with $2$ domain-walls.
	
	A complete basis in this multiplet is given by
	\begin{eqnarray}
	\ket{j,l}:=\ket{\cdots\uparrow\underbrace{\uparrow}_j \underbrace{\downarrow}_{j+1} \downarrow\cdots \downarrow 
		\underbrace{\downarrow}_{j+l} \underbrace{\uparrow}_{j+l+1}
		\uparrow \cdots}
	\end{eqnarray}
	Denoting the ground state energy of $\mathcal{H}$ (at first order in perturbation theory) with $E_0$, and defining $\ket{j,l\leq0}\equiv0$, we have
	\begin{equation}
	\frac{H-E_0}{J}\ket{j,l} =
	\left(1 - 2\lambda_{\text{AF}} + h_z l +\delta_{l,1} \lambda_{\text{AF}}\right) \ket{j,l} \\
	- \frac{\lambda_{S}}{2} \delta_{l,1} \left( \ket{j+1,1} + \ket{j-1,1} \right) + T\ket{j,l}
	\end{equation}
	with $T$, depending on the terms added to $\mathcal{H}$, being
	\begin{description}
		\item[$J \lambda_A \sum_j \left(S^x_i S^x_{j+1} - S^y_j S^y_{j+1}\right)$]
		\begin{equation}
		T_1\ket{j,l} =
		\frac{\lambda_{A}}{2}
		\left( \ket{j+2,l-2} + \ket{j-2,l+2}
		+  \ket{j,l+2} + \ket{j,l-2} \right)
		\end{equation}			
		\item[$J \lambda_{xy} \sum_j (-1)^j\left(S^x_i S^y_{i+1} + S^y_i S^x_{i+1}\right)$]
		\begin{equation}
		T_2\ket{j,l} =
		\frac{\lambda_{xy}i}{2} \left[
		(-1)^j \left( \ket{j+2,l-2} - \ket{j-2,l+2} \right)
		- (-1)^{j+l} \left( \ket{j,l+2} - \ket{j,l-2} \right)
		\right]
		\end{equation}
		
		\item[$J \lambda_{yz}\sum_j (-1)^j\left(S^y_i S^z_{i+1} + S^z_i S^y_{i+1}\right)$]
		\begin{equation}
		T_3\ket{j,l} =
		\frac{\lambda_{yz}i}{2} \left[
		(-1)^j \left( \ket{j+1,l-1} + \ket{j-1,l+1} \right)
		+ (-1)^{j+l} \left( \ket{j,l+1} + \ket{j,l-1} \right)
		\right]
		\end{equation}
	\end{description}
	
	The shape of the DW continuum is set by the form of $T$, while the other couplings result in an effective nearest-neighbour interaction between a pair of DWs.
	
	To understand the shape of the continuum for the different terms, we move to the momentum eigensates: (setting $\tilde c=1$)
	\begin{equation}
	\Ket{k,l} = \frac{1}{\sqrt{L}} \sum_j e^{ikj}\ket{j,l}
	\end{equation}
	Rewriting the $T$ matrix elements in this basis, we obtain respectively
	
	\begin{equation}
	T_1\ket{k,l} =
	\frac{\lambda_{A}}{2}
	\left[ \left(e^{-2ik}+1\right)\ket{ k,l-2} + \left(e^{2ik}+1\right)\ket{ k,l+2}
	\right]
	\end{equation}			
	
	\begin{equation}
	T_2\ket{ k,l} =
	\frac{\lambda_{xy}i}{2} \left[
	\left(e^{-2ik}+e^{i\pi l}\right)\ket{{k+\pi},l-2}-
	\left(e^{2ik}+e^{i\pi l}\right)\ket{{k+\pi},l+2}
	\right]
	\end{equation}
	
	\begin{equation}
	T_3\ket{ k,l} =
	\frac{\lambda_{yz}i}{2} \left[
	\left(e^{-ik}+e^{i\pi l}\right)\ket{{k+\pi},l-1}+
	\left(e^{ik}+e^{i\pi l}\right)\ket{{k+\pi},l+1}
	\right]
	\end{equation}
	
	The presence of a term $e^{2ik}$ in $T_1$ and $T_2$ would produce a DW continuum with periodicity $\pi/\tilde c$, i.e. half the periodicity observed experimentally.
	Instead the last term correctly produces a continuum with periodicity $2\pi/\tilde c$.
	Henceforth it is reasonable that $\lambda_A$ and $\lambda_{xy}$ are negligible w.r.t. $\lambda_{yz}$.
	
	Note that the mixing between the $k$ and $k+\pi$ sector is not enough to produce a Brillouin halving visible through the DSF. In fact, that would require the states $\ket{ k,l=1}$ and $\ket{{k+\pi},l=1}$ to belong to the same connected component of the Hilbert space, i.e. $\ket{{k+\pi},l=1}$ can be obtained by applying  the Hamiltonian on $\ket{ k,l=1}$ some arbitrary number of times.
	
	\section{Glide-symmetry counting}
	
	In this section  we work directly with the Hamiltonian $\mathcal{H}$ and re-derive the same results presented in the main text using the unfolding unitary transformation. Specifically, we rederive the enlarged BZ size probed by the DSF and the argument for QP breakdown (in the next section).
	
	Before diving into the derivation note that, in general, the origin of the correspondence between the two approaches lies in the following. A given eigenstate of $\mathcal{H}$ $\ket{n}$ will be an eigenstate of $G$, viz. $G\ket{n}=e^{i\varphi}\ket{n}$. As  $G = T_{\tilde c} \left( T_{\tilde c}^\dag U T_{\tilde c}\right) U = U T_{\tilde c} U$, the corresponding eigenstate $\ket{\tilde n}=U\ket{n}$ of $\tilde H$, satisfies $T_{\tilde c}\ket{\tilde n} = e^{i\varphi} \ket{\tilde n}$. {Therefore the glide eigenvalue is $e^{i\varphi}$ with $\varphi = \tilde k \tilde c$, where we use the same notation for the eigenvalues  $e^{i\tilde{k}\tilde{c}}$ of $T_{\tilde c}$ as in the main text.}
	
	We now turn to the derivation, First, we re-express the DSF as
	\begin{equation}
	\mathcal{S}^{\alpha,\beta}(\omega,k) = \sum_{n\neq m} \Braket{m|S^\alpha(-k)|n}\Braket{n|S^\beta(k)|m}\delta(\omega_n - \omega_m - \omega) P_m,
	\end{equation}
	where $\omega_n$ is the energy of $\ket{n}$ and $P_m\varpropto e^{-\omega_m/T}$ is the thermal probability (Boltzmann weight) associated to the eigenstate $\ket{m}$.
	
	In order to understand which $\ket{n}$ contributes to this sum, i.e. yield $\braket{n|S^\beta(k)|0}, \braket{n|S^\alpha(k)|0} \neq 0$, we insert the identity $G^\dag G=\mathds{1}$, obtaining
	\begin{equation}
	\braket{n|S^\beta(k)|m} = \braket{n|S^\beta(k)|m} \exp \left[i\left(\varphi_m + k \tilde c + \delta_\beta  - \varphi_n\right)\right]
	\end{equation}
	and similarly for $\braket{n|S^\alpha(k)|m}$. Requiring these to be non-zero yields
	\begin{eqnarray}
	\varphi_n - \varphi_m = k \tilde c + \delta_\alpha \,\,(\text{mod}\,2\pi) ,
	\qquad
	\delta_\alpha = \delta_\beta. 
	\end{eqnarray}
	In this way, $\mathcal{S}^{\alpha,\beta}(\omega,k)$ will probe the transitions with $\Delta\varphi = k \tilde c + \delta_\alpha\,\,(\text{mod}\,2\pi)$, hence to probe the same transition at two different $k$, it has to be $\Delta k = 2\pi/\tilde c$, i.e. the periodicity of the DSF will generally be $2\pi/\tilde c$, in agreement with the main text.
	
	\section{Comparison of different ways of constructing 2QP continuum}
	
	\begin{figure}
		\begin{center}
			\includegraphics[width=0.95\linewidth]{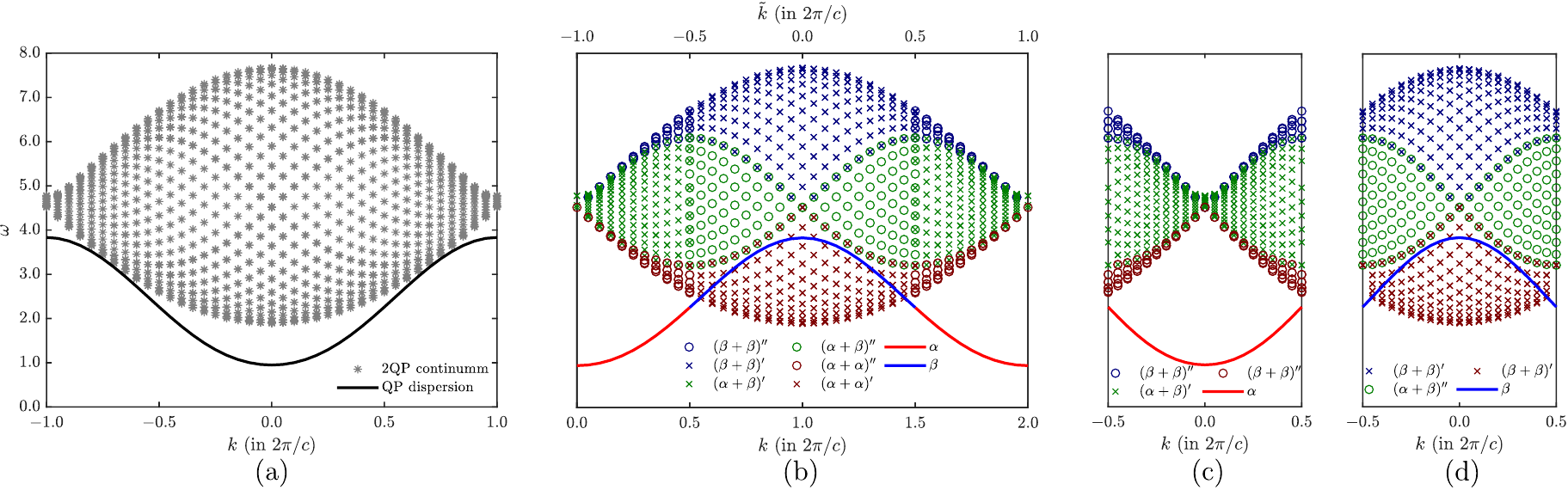}
		\end{center}
		\caption{
			(a) Single-quasiparticle dispersion $\omega_{\text{QP}}(k)$ (solid line) and the derived 2QP-continuum at $B=9$ T, as interpolated from INS data~\cite{PhysRevB.90.014418}.  Ignoring the the staggered term in $\mathcal{H}$, the 2QP continuum would be in the region denoted by grey asterisks. (b) The continuum in the $(\tilde{k},\omega)$ plane computed taking into account the glide symmetry. (c-d) Energy dispersion in the ``proper'' BZ of the material (i.e. going from $k=-\pi/(2\tilde c)$ to $k=-\pi/(2\tilde c)$). In this setting we distinguish two QP bands: $\alpha$ and $\beta$ (red and blue curves), crossing linearly at $k=\pi/(2\tilde{c})$. They give rise to different continua distinguished by (i) the 2 QP species ($\alpha+\alpha$ in red, $\alpha+\beta$ in green, and $\beta+\beta$ in blue) and (ii) whether the sum of the two momenta lies in the first ($\times$) or second ($\circ$) BZ. For clarity, we separated into two sub-panels  the 3 continua which can couple to the $\alpha$ band (c) and the 3 which can couple to the $\beta$ band (d). Note that only the $\beta$ quasiparticles can kinematically decay.
		}
		\label{app:fig:continuum}
	\end{figure}
	
	In this section we stress the differences in the position of the 2QP continuum obtained in two ways: (i) by {\it incorrectly} counting momentum while ignoring the presence of a staggered term and (ii) {\it correctly}, by counting glide-symmetry eigenvalues.
	
	In the first case, the 1QP band $\omega(k)$ can be obtained ``as is'' from the DSF. 2QP continuum states are then states with two such excitations at momenta $k_1$ and $k_2$. In this way the quantum number of the 2QP state is $k=k_1+k_2$, $\omega = \omega(k_1) + \omega(k_2)$. The result obtained in this way will resemble those shown in Fig.~\ref{app:fig:continuum}(a).
	
	Instead, in the second case, we first recognize that the dominant components of the DSF observed in INS are $aa$ and $cc$. Then the 1QP dispersion in the glide eigenvalue ($\tilde k$) is obtained by shifting the $\omega(k)$ band in the DSF by $k\mapsto \tilde k = k + \pi/\tilde c$ as in Fig.~\ref{app:fig:continuum}(a). A 2QP state is one formed by two QP with glide-number $\tilde k_1$ and $\tilde k_2$. Neglecting interactions between the individual QPs, this 2QP state has energy $\omega(\tilde k_1,\tilde k_2)=\omega(\tilde k_1)+\omega(\tilde k_2)$, and  glide eigenvalue $\tilde k(\tilde k_1,\tilde k_2) =  \tilde k_1 + \tilde k_2$. This last relationship can be most conveniently obtained by performing the $U$ transformation, exploiting the additivity of momentum and, finally, transforming back with $U$. The position of the continuum will be qualitatively similar to Fig.~\ref{app:fig:continuum}(b).
	
	Finally, we outline an equivalent way of describing the QP breakdown in the glide-invariant system. If we were to ignore the glide symmetry, we would employ a BZ of full length $\pi/\tilde{c}$. The QP dispersion in the halved BZ is obtained by folding the QP dispersion as seen e.g. by INS. Due to the folding, in the smaller BZ there will appear to be $2$ separate bands, which we denote by $\alpha$ and $\beta$ (Fig.~\ref{app:fig:continuum}(c)). At $k=\pi/(2\tilde{c})$ the two bands cross linearly. While linear crossings are generally unstable in 1D, this linear crossing is protected by glide symmetry (for an example where this is worked out explicitly see Ref.~\cite{Parameswaran2019}). Since there are two bands, we can construct $3$ different continua depending on the species of the two QP: $\alpha+\alpha$, $\alpha+\beta$, $\beta+\beta$. Furthermore, for each pair of QP species, we need to distinguish two sub-cases, depending on whether the sum of the two QP momenta lies in the first or second BZ, which we denote by $(\cdots)'$ and $(\cdots)''$ respectively.  In this picture, the decay of a QP into the $2$-QP continua is not only constrained by $k$, but also by the glide eigenvalue ({Note that this step is unnecessary if one directly considers $\tilde k$ , which is in $1$-to-$1$ correspondence with the glide eigenvalues}). Imposing the glide eigenvalue constraints, we find that symmetry-allowed decay channels are as follows (see also Fig.~\ref{app:fig:continuum}(c))
	\begin{equation}
	\alpha \to \left\{
	\begin{array}{c}
	(\alpha + \alpha)''\\
	(\alpha + \beta)'\\
	(\beta + \beta)''
	\end{array}
	\right.
	\qquad
	\beta \to \left\{
	\begin{array}{c}
	(\alpha + \alpha)'\\
	(\alpha + \beta)''\\
	(\beta + \beta)'
	\end{array}
	\right.
	\end{equation}

	\section{DSF for off-axis wave-vectors and ``shadow mode''}
	
	\begin{figure}
		\begin{center}
			\includegraphics[width=0.5\linewidth]{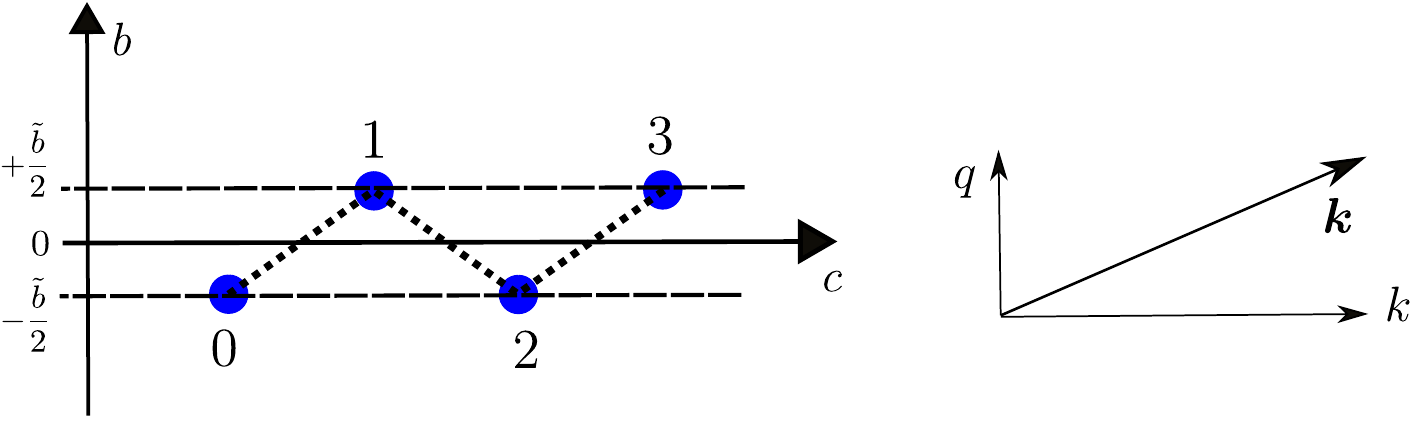}
		\end{center}
		\caption{Geometry of the chain with even/odd sites respectively in position $\mp\tilde{b}/2$.}
		\label{app:fig:misaligned}
	\end{figure}
	
	{In Fig.~3
		, we reported the experimental measurement of the DSF. As noted in the main text, the experimental DSF in Fig.~3 is measured through INS where the momentum transfer of the neutrons is not aligned in the $c$-direction, but also has a component in the $b$-direction. We mentioned in the main text that, due to the buckled chain geometry, this results in the INS data showing a linear superposition of $\mathcal{S}(\omega, k)$ and $\mathcal{S}(\omega, k+\pi/\tilde{c})$.}
	
	{In this section we justify our statement by computing} the DSF $\mathcal{S}(\omega, \bk)$ when the wavevector $\bk$ has a non-zero component perpendicular to the $c$-axis (see Fig.~\ref{app:fig:misaligned}): $q$ in the $b$ direction and $q'$ in the $a$ direction, so that $\bk = (q',q,k)$. We will show that the DSF can be expressed as a linear combination of the $1d$ DSF $\mathcal{S}(\omega,k)$ and $\mathcal{S}(\omega,k+\pi/\tilde{c})$ --- as well as other $1$D terms, that are however small in the case of $\mathrm{CoNb_2O_6}$. The $\mathcal{S}(\omega,k+\pi/\tilde{c})$ component gives rise to the ``shadow mode'' discussed in Ref.~\cite{PhysRevB.90.014418}.
	
	The DSF in this more general case is
	\begin{align}
	{\mathcal{S}}^{\alpha\beta}(\omega, \bk) \equiv& \frac{1}{L} \int_{-\infty}^\infty d t \sum_{l,m} e^{i\bk(\boldsymbol{r}_l - \boldsymbol{r}_{m})} e^{i\omega t} \langle S^\alpha_ l(t) S^\beta_{m}\rangle \\
	=& \frac{1}{L} \int_{-\infty}^\infty d t \sum_{l,m} e^{ik(l-m)\tilde{c}} e^{i\omega t} f(l,m) \langle S^\alpha_ l(t) S^\beta_{m}\rangle
	\end{align}
	with
	\begin{equation}
	f(l,m)= \left\{\begin{array}{cc} e^{iq\tilde b} & \text{ if } l \text{ is odd and } m \text{ is even} ,\\
	e^{-iq\tilde b} & \text{if } l \text{ is even and } m \text{ is odd} ,\\
	1& \text{otherwise}, \end{array} \right.
	\end{equation}
	as even(odd) sites have $b$ coordinate of $-\tilde{b}/2$ ($\tilde{b}/2$).
	Explicitly splitting the sum for even and odd $m$, and re-expressing sums over $l$ using
	\begin{equation}
	\sum_l f(l,m) (\cdots) = \sum_l \left[ f(0,m) \frac{1+e^{i\pi l}}{2} + f(1,m) \frac{1-e^{i\pi l}}{2} \right] (\cdots)
	\end{equation}
	we obtain
	\begin{align}
	\begin{split}
	\mathcal{S}^{\alpha\beta}(\omega, \bk) =& \mathcal{S}^{\alpha\beta}(\omega, k) \frac{1 + \cos(q\tilde{b})}{2}
	+ \mathcal{S}^{\alpha\beta}(\omega, k+\pi/\tilde{c}) \frac{1 - \cos(q\tilde{b})}{2}\\
	+& \left[{\mathcal{S}}^{\alpha\beta}_A (\omega, k) - {\mathcal{S}}^{\alpha\beta}_A (\omega, k+\pi/\tilde{c}) \right] \sin(q\tilde{b})
	\end{split}
	\end{align}
	where we introduced the function ${\mathcal{S}}^{\alpha\beta}_A = i\left({\mathcal{S}}^{\alpha\beta}_0 - {\mathcal{S}}^{\alpha\beta}_1\right)$ defined in terms of
	\begin{align}
	{\mathcal{S}}^{\alpha\beta}_0 (\omega, \bk) \equiv& \frac{1}{L} \int_{-\infty}^\infty d t \sum_{l} \sum_{\text{even } m} e^{ik(l -m)\tilde{c}} e^{i\omega t} \langle S^\alpha_ l(t) S^\beta_{m}\rangle \\
	{\mathcal{S}}^{\alpha\beta}_1 (\omega, \bk) \equiv& \frac{1}{L} \int_{-\infty}^\infty d t \sum_{l} \sum_{\text{odd } m} e^{ik(l -m)\tilde{c}} e^{i\omega t} \langle S^\alpha_ l(t) S^\beta_{m}\rangle.
	\end{align}
	{Splitting ${\mathcal{S}}^{\alpha\beta}_0$ and ${\mathcal{S}}^{\alpha\beta}_1$ into two separate terms depending on the parity of $l$, and exploiting translation invariance by two sites, one can explicitly verify that the combination $\left[{\mathcal{S}}^{\alpha\beta}_A (\omega, k) - {\mathcal{S}}^{\alpha\beta}_A (\omega, k+\pi/\tilde{c}) \right]$ is real, as expected.}
	
	Finally, note that ${\mathcal{S}}^{\alpha\beta}_A$ would vanish if no staggered term were present. For $\mathrm{CoNb_2O_6}$, as $\lambda_{yz}$ is a weak correction to the uniform component of $\mathcal{H}$, ${\mathcal{S}}^{\alpha\beta}_A$ will be small. {The dominant effect is then the one mentioned in the main text, i.e. INS data shows a superposition of $\mathcal{S}(\omega, k)$ and $\mathcal{S}(\omega, k+\pi/\tilde{c})$}.
	
	\section{Convergence analysis of tDMRG simulations}
	In Fig.~\ref{app:fig:h_0_conv} we verify the convergence of our simulations for $h_y=0$, and in Fig.~\ref{app:fig:B_2.5_conv} for $B=2.5$ T. In these cases, we did not employ an hard-cutoff $\chi$ since the entanglement growth is very mild.
	The convergence analysis for the data in Fig.3
	is performed in Fig.~\ref{app:fig:para_convergence}.
	The larger $\eta$ used for the simulation in non-zero field are due to the fact that the dynamics in this case produces entanglement more rapidly, therefore $t_{max}$ for $h_y\neq0$ is shorter than $t_{max}$ at $h_y=0$.
	
	\begin{figure}
		\begin{center}
			\includegraphics[width=0.5\linewidth]{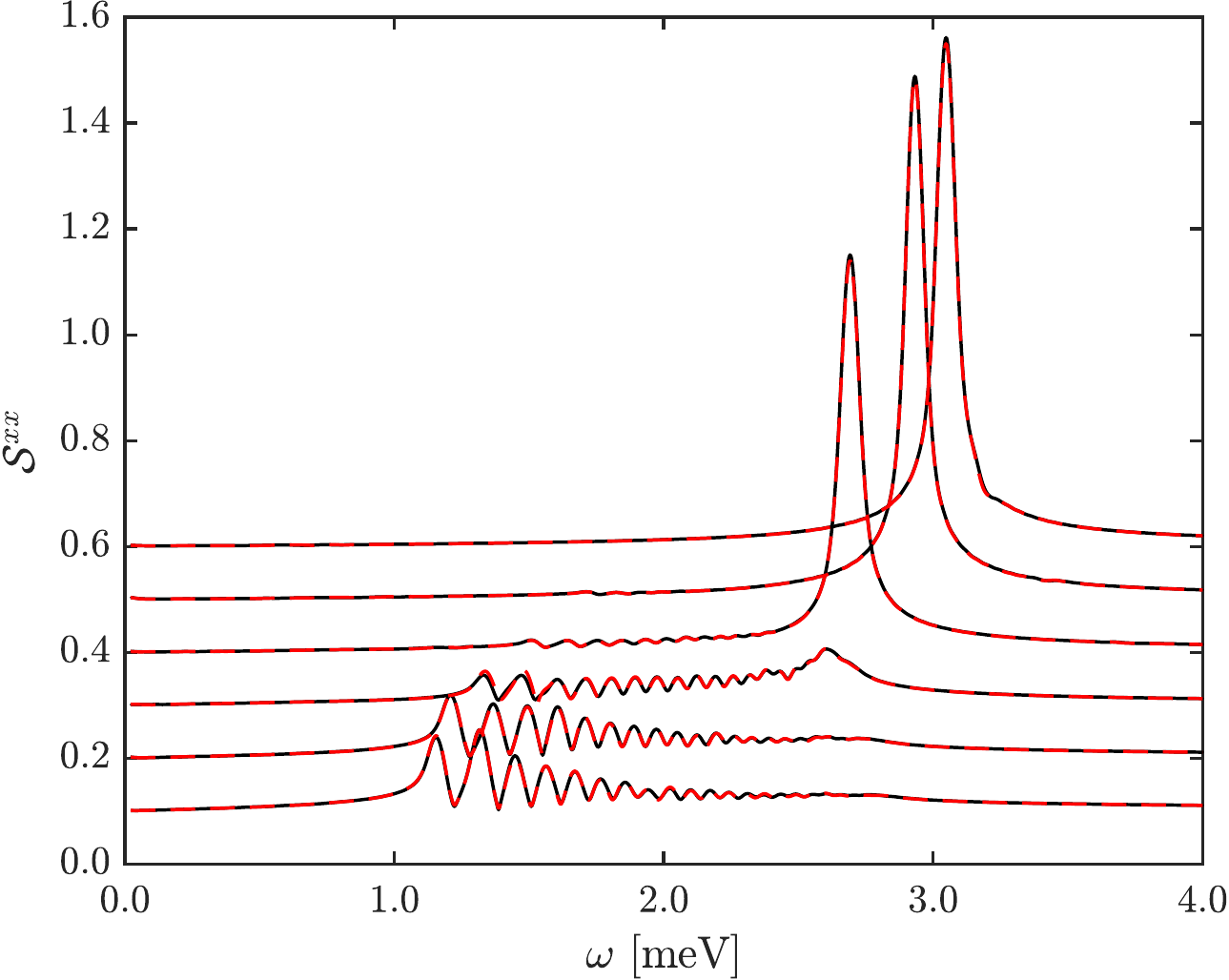}
		\end{center}
		
		\caption{Convergence analysis for the simulations at $h_y=0$. {Lines denotes cuts of $\mathcal{S}^{xx}(\omega,k)$ at linearly spaced values of $k$, going from $0$ to $\pi/\tilde{c}$. For clarity, as $k$ increases the curves are vertically displaced.} The black curve is obtained with tDMRG parameters $\varepsilon=10^{-10}$ and $\delta t = 5\cdot10^{-3}$. Instead, the parameters of the red curve are the same as in Fig.~2
			. In both cases, $t_{max}=500/J$ and $\eta=J/125$. The positions of the maxima display a good convergence, especially around $k=0$ and $\pi/\tilde{c}$.}
		\label{app:fig:h_0_conv}
	\end{figure}
	
	\begin{figure}
		\begin{center}
			\includegraphics[width=0.5\linewidth]{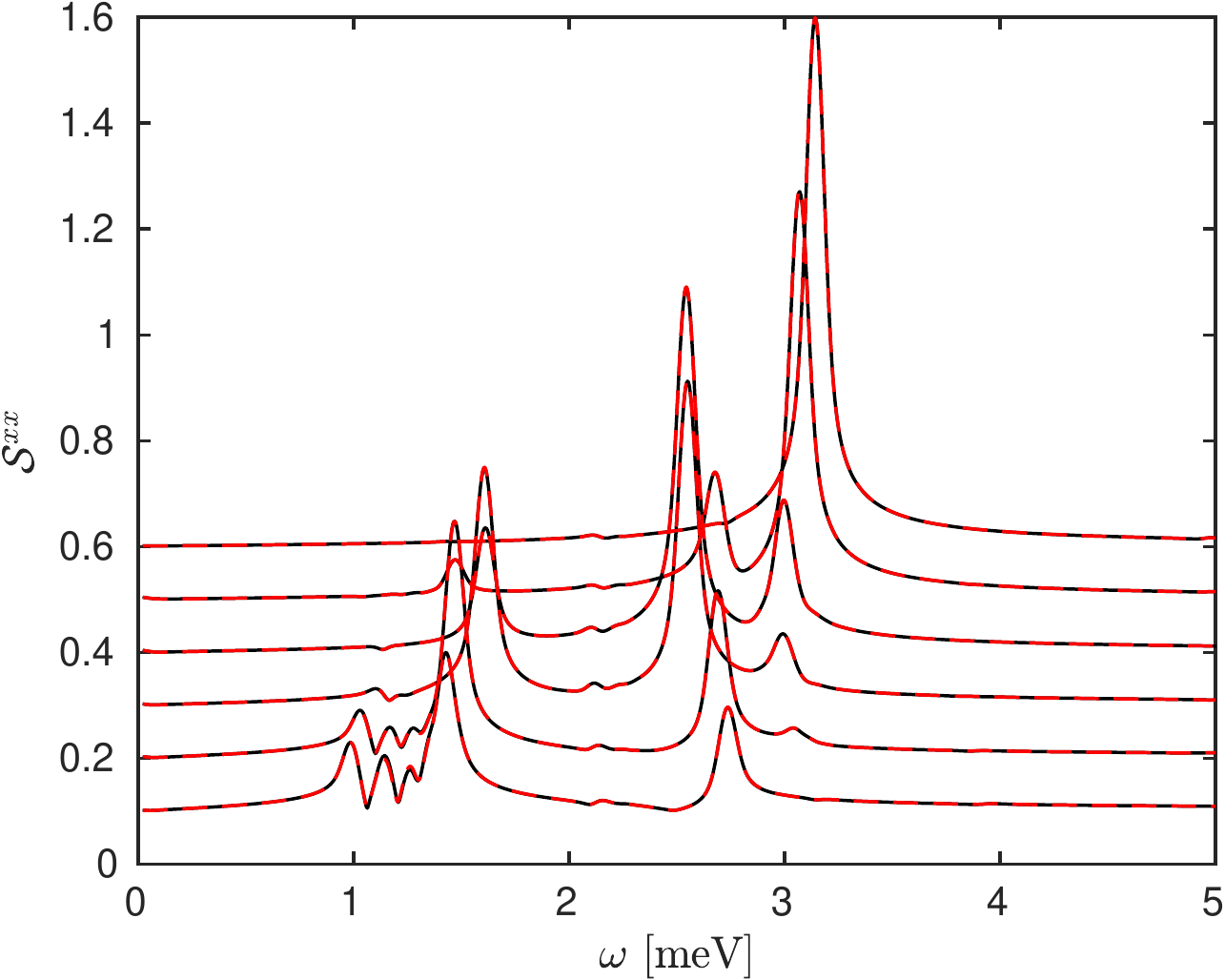}
		\end{center}
		
		\caption{Convergence analysis for the simulations at $B=2.5$ T. {Lines denotes cuts of $\mathcal{S}^{xx}(\omega,k)$ at linearly spaced values of $k$, going from $0$ to $\pi/\tilde{c}$. For clarity, as $k$ increases the curves are vertically displaced.} The black curve is obtained with tDMRG parameters $\varepsilon=10^{-10}$ and $\delta t = 5\cdot10^{-3}$. Instead, the parameters of the red curve are the same as in Fig.~4
			. In both cases, $t_{max}=400/J$ and $\eta=J/100$}
		\label{app:fig:B_2.5_conv}
	\end{figure}
	
	\begin{figure}
		\begin{center}
			\includegraphics[width=0.32\linewidth]{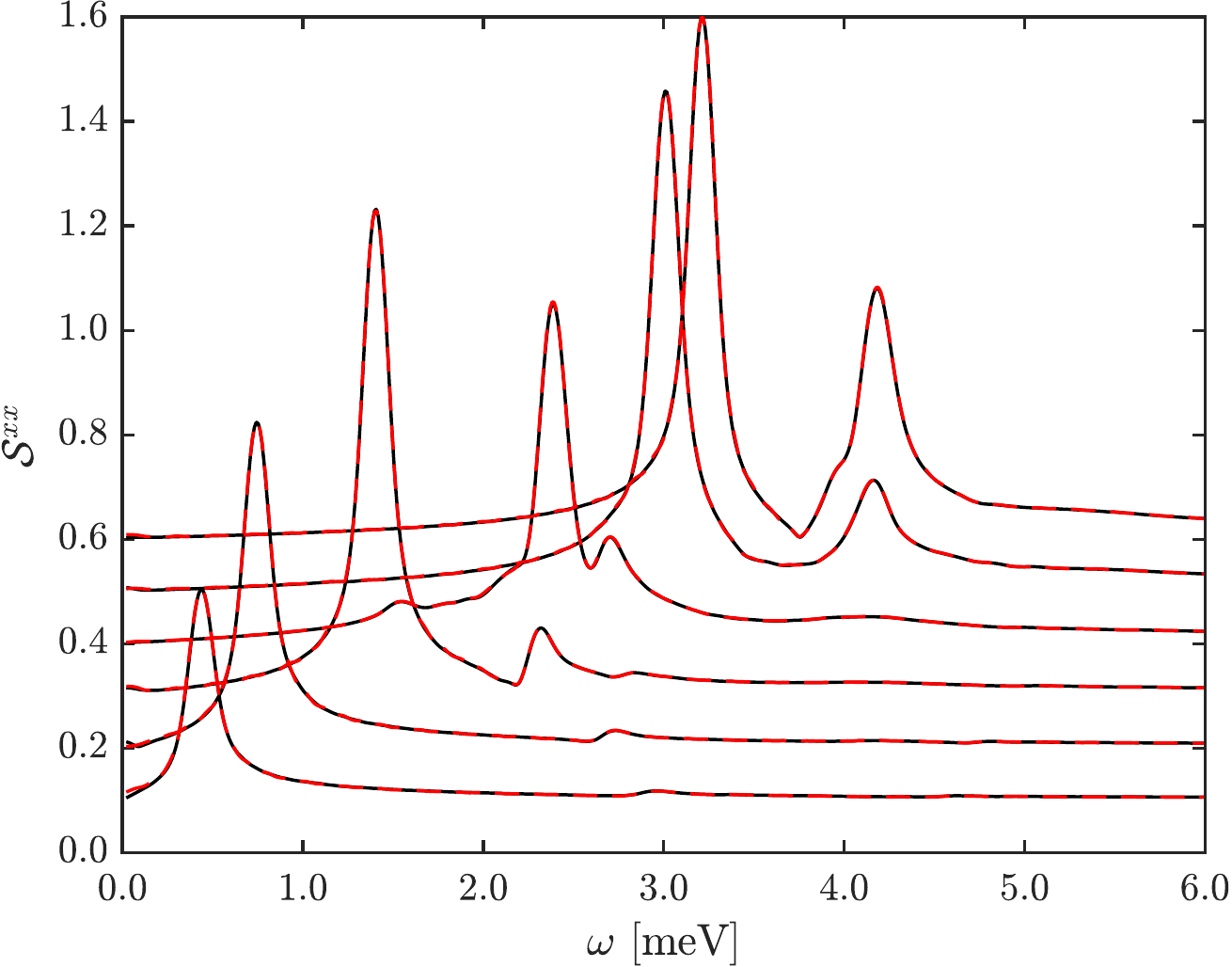}
			\includegraphics[width=0.32\linewidth]{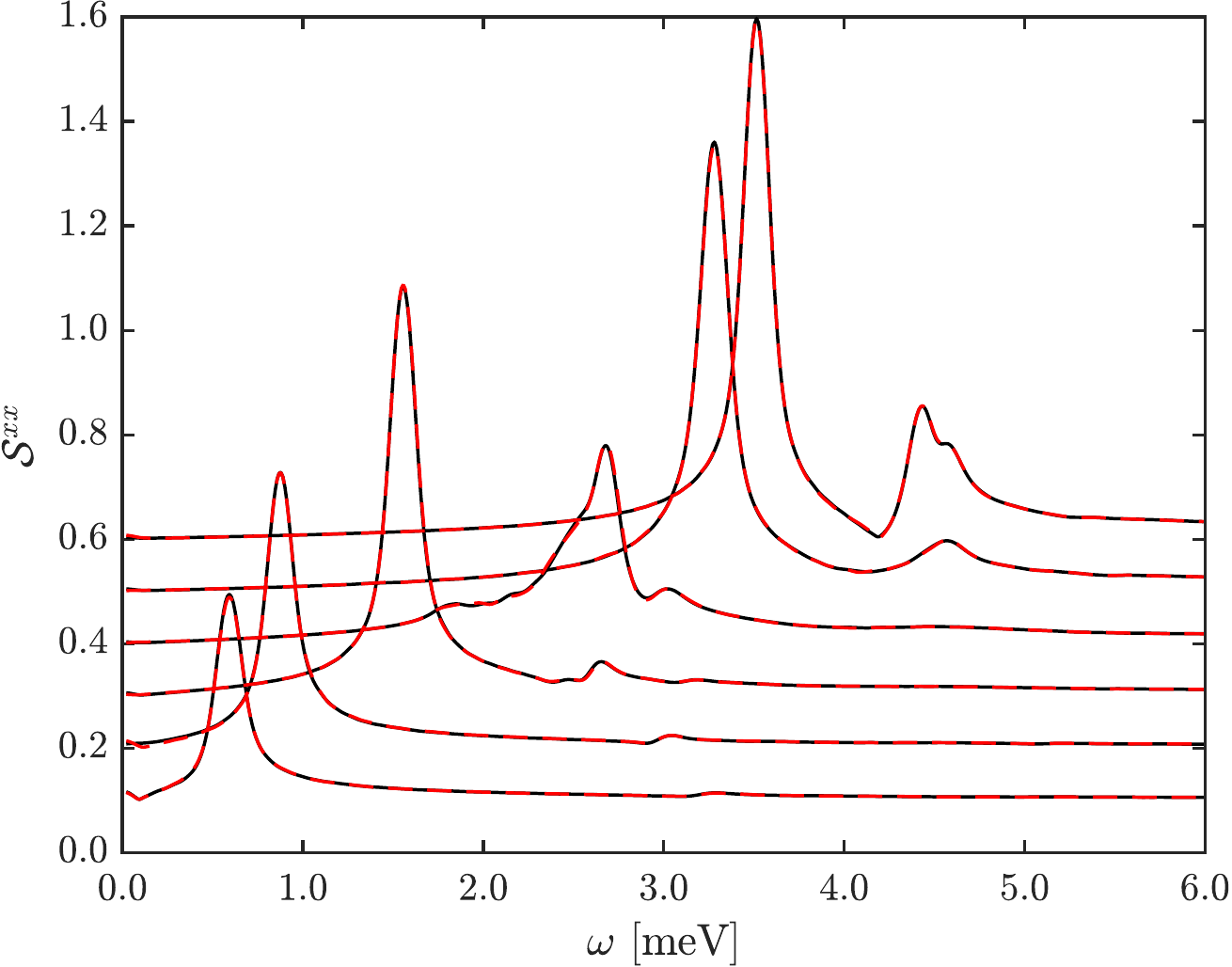}
			\includegraphics[width=0.32\linewidth]{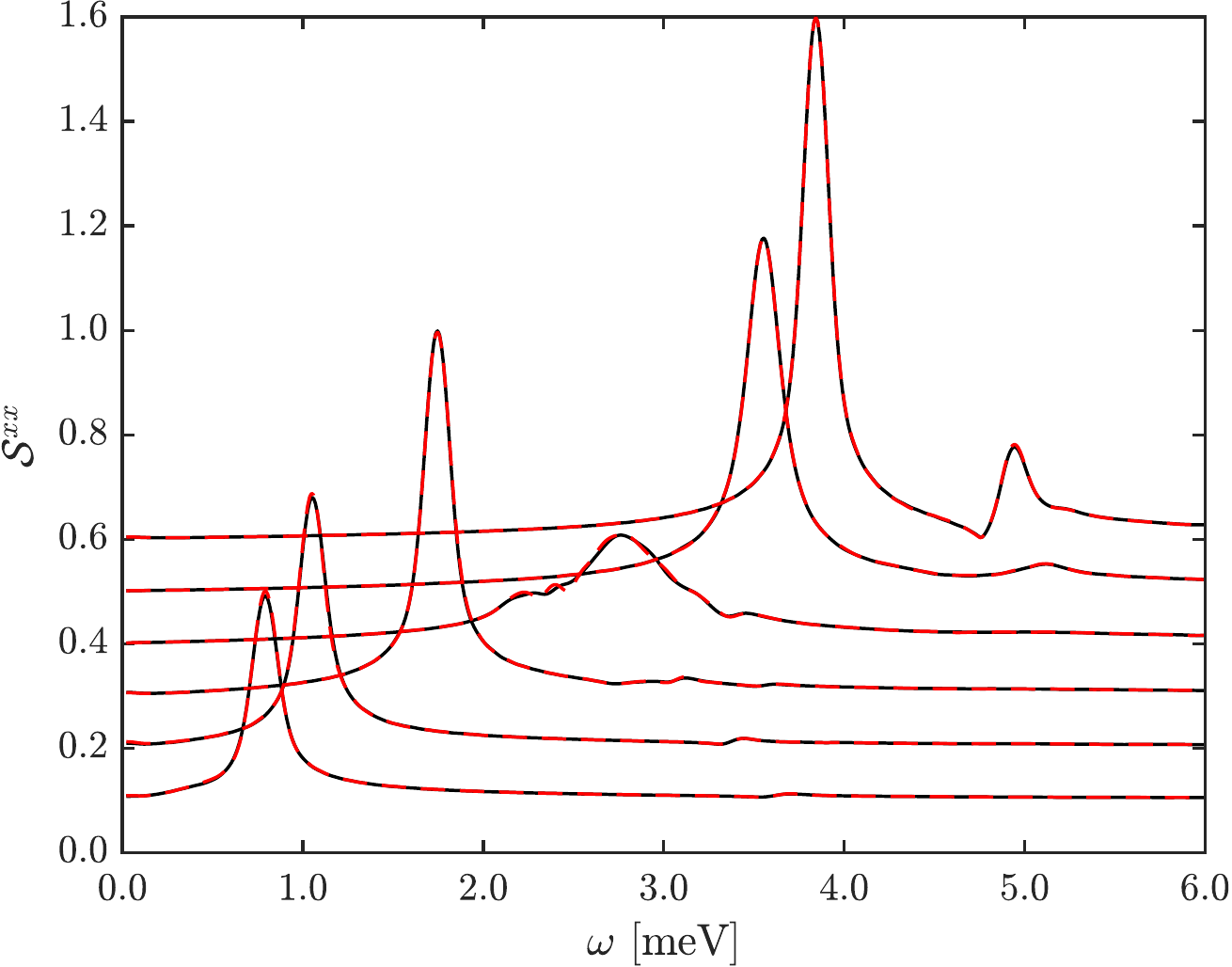}
		\end{center}
		
		\caption{Convergence analysis for the simulations at (from left to right) $B=7\mbox{T},\,8\mbox{T},\,9\mbox{T}$. Lines denotes cuts of $\mathcal{S}^{xx}(\omega,k)$ at linearly spaced values of $k$, going from $0$ to $\pi/\tilde{c}$. For clarity, as $k$ increases the curves are vertically displaced. The dashed red line is obtained with the tDMRG parameters of  Fig.~3
			; the solid black line, instead, with $\varepsilon=10^{-9}$, $\chi_{max}=200$ and $\delta t = 0.01$. In both cases, $t_{max}=200/J$ and $\eta=J/125$.}
		\label{app:fig:para_convergence}
	\end{figure}

	\section{Fit of $g_b$}
	
	\begin{figure}
		\begin{center}
			\includegraphics[width=0.5\linewidth]{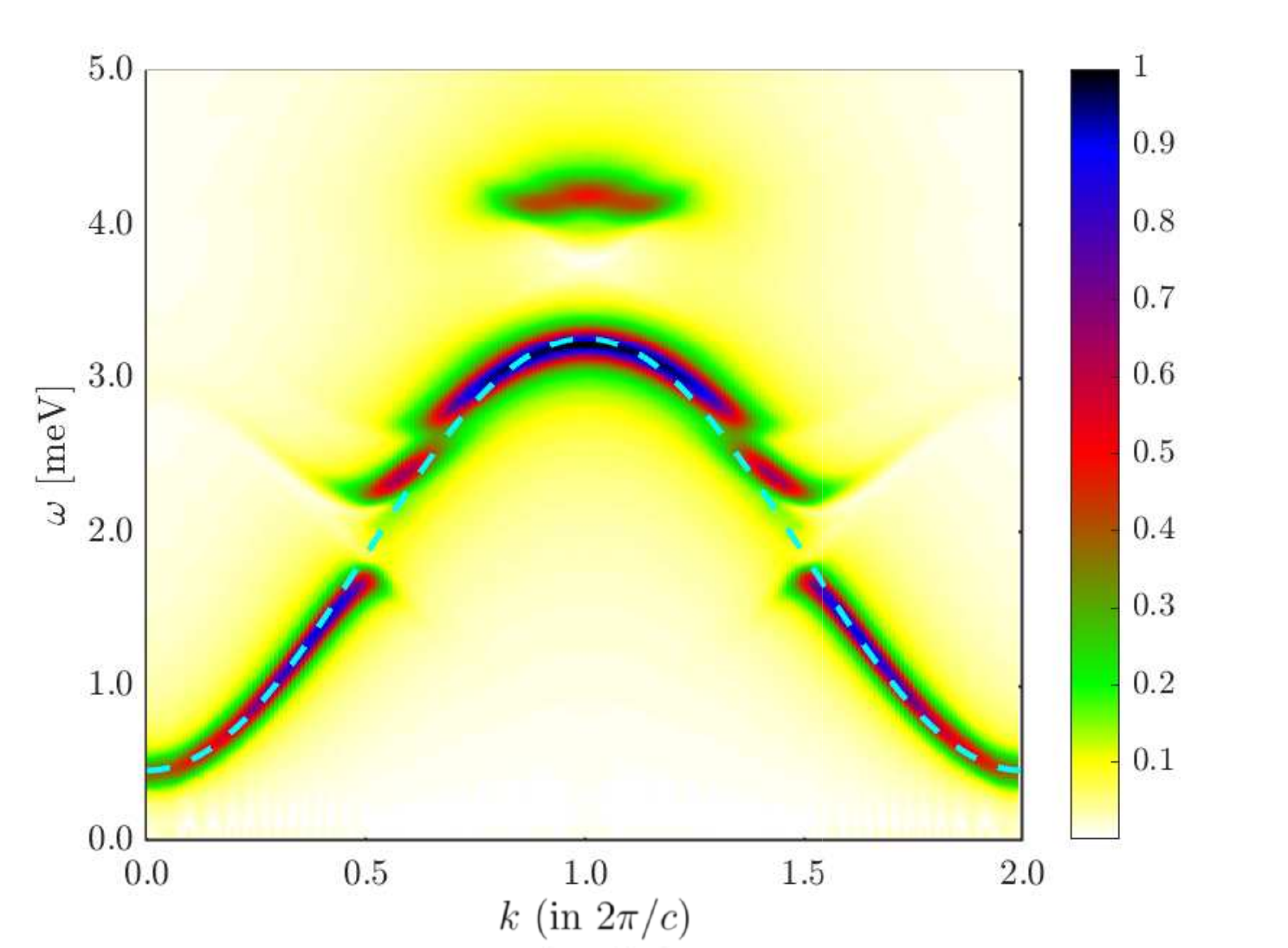}
		\end{center}
		\caption{The colorplot shows $\mathcal{S}^{xx}(\omega,k)$ at $B=7\mbox{T}$, which is the same as in Fig.~3
			. The dashed line is the single-particle dispersion relation extracted from fits to the experimental INS data and extrapolated to the case of isolated chains with no interchain couplings~\cite{PhysRevB.90.014418}.}
		\label{app:fig:g_fit}
	\end{figure}
	
	In order to determine the $g$-factor $g_b$, we consider the position of the maxima in the DSF for $\mathcal{H}$ and fit it to the parametrization of the 1QP band at $B=7\mbox{T}$ reported in~\cite{PhysRevB.90.014418}. Again, we exploit ED diagonalization on small system sizes. Therefore, to reduce finite-size effects, we consider only the lower part of the band (i.e. the one lying outside the continuum), finding $g_b\simeq3.100$. Fig.~\ref{app:fig:g_fit} shows the comparison of the tDMRG simulation with the experimental parametrization of the 1QP band.

	\section{Unfolding for other Non-Symmorphic Symmetries}
	
	The unfolding discussion applies also to other geometries and space groups whenever a non-symmorphic symmetry is a property of the embedding of a low-dimensional layer or chain into 3D space. The possible cases are, apart from the a chain buckling across a glide plane already discussed, (i) a chain twisting around a screw axis and (ii) a layer buckling across a glide plain. Both cases can be dealt as the one treated in the main text, with the only difference of the unitary unfolding transformation $U$.

	For first case, consider an $n$-fold screw axis, inducing the symmetry transformation $S=T_{\tilde c} \exp\left(-i2\pi S_j^z/ n\right)$. The unit cell of the chain will have size $c= n\tilde c$. However, taking $U = \otimes_j \exp\left(i2\pi S_j^z/ n\right)^{m_j}$ with $m_j = j$ (mod $n$) guarantees that $U^\dag \mathcal{H} U$ is invariant under $T_{\tilde c}$. In this way the BZ probed by the DSF is $2\pi/\tilde c$, i.e. $n$ times larger than the ``geometric'' BZ of size $2\pi/c$.
	
	With regard to case (ii), there are different possibilities given by the geometry of the 2D lattice. Since the unfolding transformation will depend on those, a detailed treatment is beyond the scope of this work.
	
	{
		\section{Comparison with models not featuring the domain-wall hopping term}
		
		In this section we compare the DSF obtained from the Hamiltonian in Eq.~(1)
		, with the DSF obtained from different Hamiltonians not featuring the domain-wall hopping term in Eq.~(6).
		
		We first compare the the results in Fig.~3
		with the DSF computed from the model in Ref.~\cite{PhysRevB.90.174406}, which has been fine-tuned to fit the INS data reported in the aforementioned figure. Looking at Fig.~\ref{fig:yz_0_para}, we see that the quasiparticle breakdown in this model manifests itself only as a small brightness reduction of the quasiparticle mode. Instead both the INS data and our model show a much more pronounced brightness reduction, with the quasiparticle mode almost becoming invisible on the scale of the plot.
		
		\begin{figure}
			\centering
			\includegraphics[width=\linewidth]{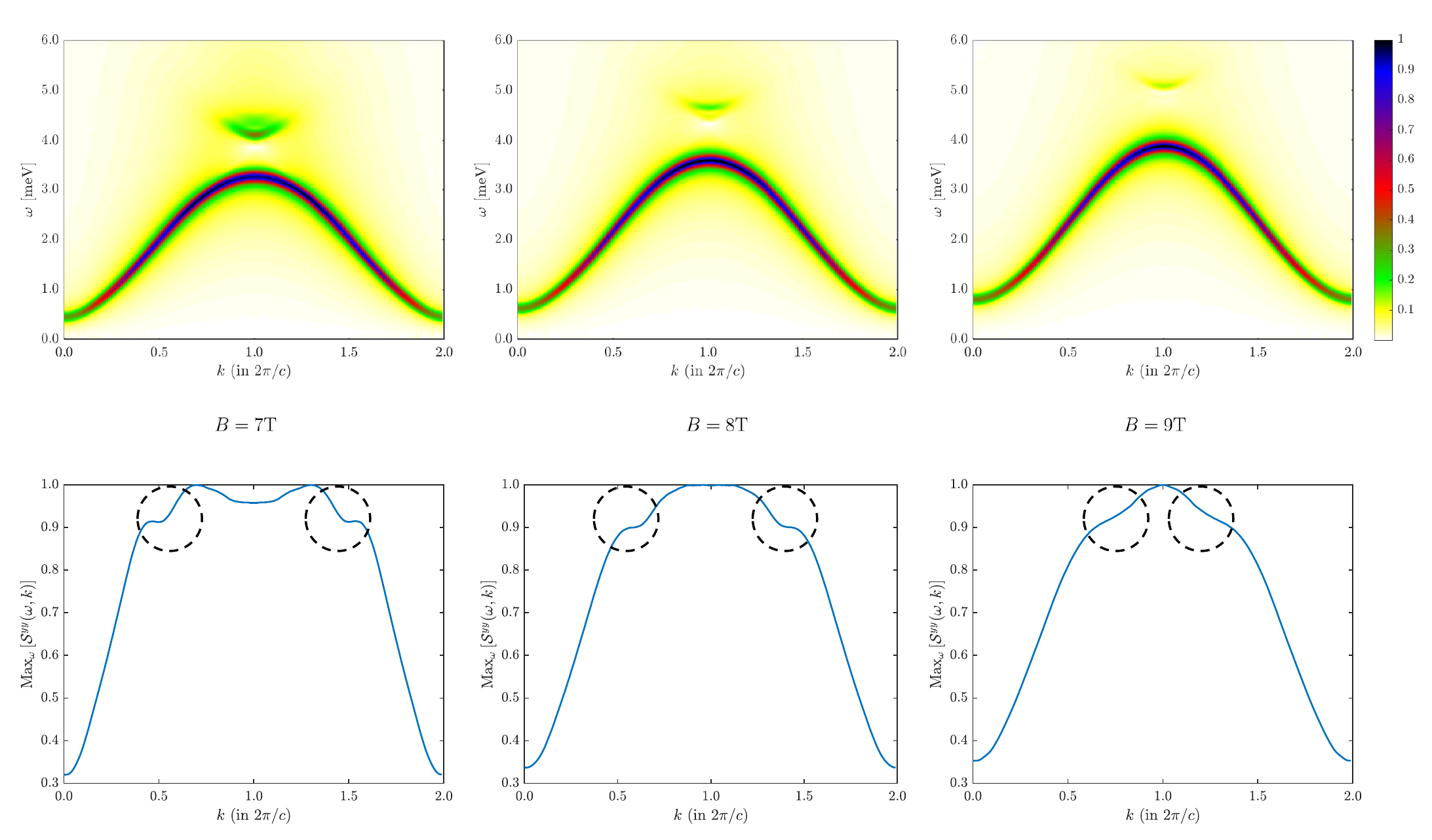}
			\caption{ DSF $\mathcal{S}^{yy}(\omega,k)$ for the model in Ref.~\cite{PhysRevB.90.174406} --- note that the $y$-direction in Ref.~\cite{PhysRevB.90.174406} is the direction perpendicular to both the ferromagnetic axis and the external field, playing the same role as the $x$-direction in our work. The lower panels report $\text{Max}_{\omega} \mathcal{S}(\omega,k)$ as a function of $k$. In this model quasiparticle breakdown is manifest as a small reduction of the quasiparticle mode brightness, highlighted by dashed circles. tDMRG parameters: $\varepsilon=2\cdot10^{-10}$, $\chi_{max}=300$, $\delta t = 0.005$, $t_{max}=300/J$ and $\eta=J/60$. Note that there is not even a qualitative match to experiments, far less a quantitative one, in contrast to the model proposed by us.}
			\label{fig:yz_0_para}
		\end{figure}
		
		Finally, we show how Fig.~4 changes if the domain wall hopping term is removed from $\mathcal{H}$. By simply setting $\lambda_{yz}=0$ in $\mathcal{H}$, we obtain Fig.~\ref{fig:yz_0_B_25}. The result we obtain is qualitatively different from the experimental data in Fig.~4a. Instead, we obtain a DSF qualitatively similar with the zero-field spectrum in Fig.~2. Indeed, whenever $\lambda_{yz}=0$ and $h_y \neq 0$, perturbation theory about the fully magnetized state will yield the same effective Hamiltonian of Ref.~\cite{Coldea2010}. Thus, if $\lambda_{yz}=0$, we always expect to find a DSF qualitatively similar to the one in Fig.~\ref{fig:yz_0_B_25}  (see also Ref.~\cite{PhysRevB.83.020407}, which studies precisely the  $\lambda_{yz}=0$ $h_y \neq 0$ case).
		
		\begin{figure}
			\centering
			\includegraphics[width = 0.4\linewidth]{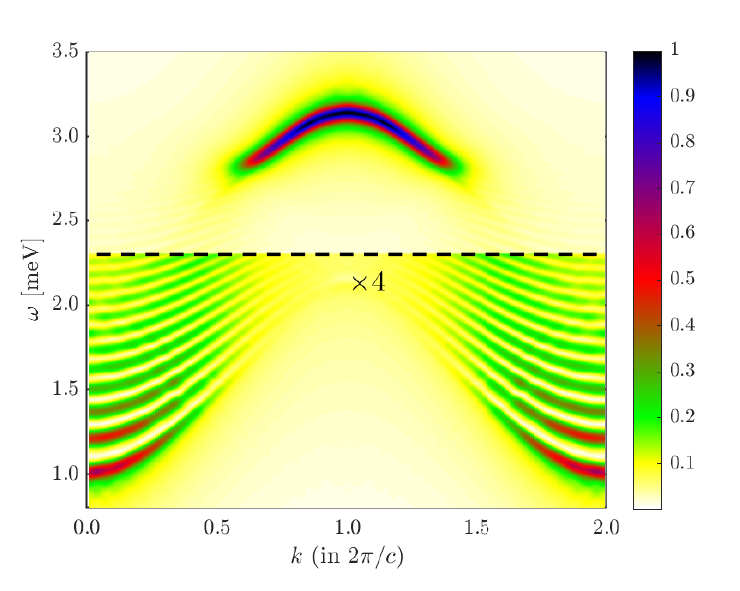}
			\caption{ DSF $\mathcal{S}^{xx}(\omega,k)$ for the Hamiltonian in Eq.~(1)
				, but $\lambda_{yz}=0$, at $B=2.5\mbox{T}$. tDMRG parameters: $\varepsilon=2\cdot10^{-11}$, $\delta t = 2.5\cdot 10^{-3}$, $t_{max}=400/J$ and $\eta=J/125$.}
			\label{fig:yz_0_B_25}
		\end{figure}

	}
	
\end{appendix}

\end{document}